\documentclass[aip,reprint,amsmath,amssymb]{revtex4-1}

\usepackage{graphicx} 
\usepackage{dcolumn}
\usepackage{bm}
\usepackage{amsmath}
\usepackage{empheq}

\draft 

\begin{document}

\title{A compact and stable incidence-plane-rotating second harmonics detector}

\author{S.~H.~Kim}
\affiliation{Center for Correlated Electron Systems, Institute for Basic Science, Seoul 08826, Republic of Korea}
\author{S.~Jung}
\author{B.~Seok}
\author{Y.~S.~Kim}
\affiliation{Center for Correlated Electron Systems, Institute for Basic Science, Seoul 08826, Republic of Korea}
\affiliation{Department of Physics and Astronomy, Seoul National University, Seoul 08826, Republic of Korea}
\author{H.~Park}
\affiliation{Department of Physics and Astronomy, Seoul National University, Seoul 08826, Republic of Korea}
\author{T.~Otsu}
\author{Y.~Kobayashi}
\affiliation{ISSP, The University of Tokyo, 5-1-5 Kashiwa-no-ha, Kashiwa, Chiba 277-8581, Japan}
\author{C.~Kim}
\affiliation{Center for Correlated Electron Systems, Institute for Basic Science, Seoul 08826, Republic of Korea}
\affiliation{Department of Physics and Astronomy, Seoul National University, Seoul 08826, Republic of Korea}
\author{Y.~Ishida}
\email[]{ishiday@issp.u-tokyo.ac.jp}
\affiliation{Center for Correlated Electron Systems, Institute for Basic Science, Seoul 08826, Republic of Korea}
\affiliation{ISSP, The University of Tokyo, 5-1-5 Kashiwa-no-ha, Kashiwa, Chiba 277-8581, Japan}

\date{\today}

\begin{abstract}
We describe a compact and stable setup for detecting the optical second harmonics, in which the incident plane rotates with respect to the sample. The setup is composed of rotating Fresnel-rhomb optics and a femtosecond ytterbium-doped fiber-laser source operating at the repetition frequency of 10~MHz. The setup including the laser source occupies an area of 1~m$^2$ and is stable so that the intensity fluctuation of the laser harmonics can be less than 0.2~\% for 4~h. We present the isotropic harmonic signal of a gold mirror of 0.5~pW and demonstrate the integrity and sensitivity of the setup. We also show the polarization-dependent six-fold pattern of the harmonics of a few-layer WSe$_2$, from which we infer the degree of local-field effects. Finally, we describe the extensibility of the setup to investigate the samples in various conditions such as cryogenic, strained, ultrafast non-equilibrium, and high magnetic fields.
\end{abstract}


\maketitle 
\draft 

\section{Introduction}
\label{sec_Intro}

Optical second harmonics can be generated when the medium is shaken with intense optical field of frequency $\omega$, $\mathbf{E}(\omega)$:~\cite{61PRL_Franken} The octave vibrionic component of the medium, namely the induced polarization $\mathbf{P}(2\omega)$, generates the second harmonics. Second harmonics generation (SHG) can be manifested profoundly when the medium is subjected to broken inversion symmetry.~\cite{Boyd} Thus, SHG has been widely used to investigate the crystallographic symmetry and underlying electronic structural phase related to non-centrosymmetry. The SHG technique is also sensitive to surface, edges, hinges and corners of the crystal,~\cite{14Sci_MoS2_Edge} where the inversion symmetry is intrinsically broken and polarization charge remains macroscopic. This aspect makes the SHG technique also useful and attractive to investigate the surface and interface phenomena~\cite{89YRShen,91Heinz} as well as the edge modes of topological materials~\cite{11PRL_Hsieh_BiSe_Aging,11PRL_Hsieh_BiSe_Dynamics} including those of higher-order topological insulators.~\cite{17Sci_HOTI_Theory,18Nature_HOTI_Photonic} Another recent trend is to artificially enhance the nonlinear response in meta-materials and meta-surfaces.~\cite{06Science_MagneticMeta,18MaterToday_metasurface,18NanoLett_SolelyChiral} 

From the beginning of the technique, the SHG measurements had been performed mostly in the normal incidence configuration. This configuration is enough to figure out the SHG signal generated by $P_x$ and $P_y$ but not by the out-of-plane polarization $P_z$. Investigation of the latter requires us to illuminate the sample with oblique (off-normal) incidence. 

\begin{figure}[htb]
\includegraphics{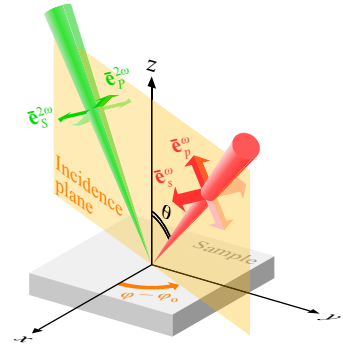}
\caption{\label{fig_IncRot} Incidence-plane-rotating SHG detection. The incident beam and the SHG signal detector rotate about the surface normal of the sample. The incident beam has to keep the \textit{p} or \textit{s} polarization during the rotation.}
\end{figure}

There are three ways to attempt rotations in the oblique incidence configuration: (1) polarization, (2) sample and (3) incidence-plane rotation methods. In the polarization rotation method,~\cite{04Science_InterfaceMagnetism,09PRB_Ogawa_LAOSTO,15NNano_WSe2_SHG_Electric_ZDXu,17NPhys_TaAs_SHG_LuOrenstein,19PRX_MoTe_SHG_NLWang} the beam and detector are spatially fixed and the incident beam polarization is rotated. Since the scattering plane is fixed with respect to the sample, this method enables access to only a limited number of the tensor components of the nonlinear optical polarizability. In the second method,~\cite{83PRL_Si_VacuumRot,93PRL_GaAs_VacuumRot,00JAP_Sato_Magnetic_Rotation,10PRL_GoldNano_SHG_Valev} the sample is rotated while the beam and detector are spatially fixed and investigations into all the tensor components become possible. However, its application to the samples in non-ambient conditions such as cryogenic, high magnetic and ultra-high vacuum is limited~\cite{83PRL_Si_VacuumRot,93PRL_GaAs_VacuumRot} because it is generally difficult to rotate the sample precisely therein. Finally, one can use the incidence-plane rotation method (Fig.~\ref{fig_IncRot}), in which the sample is fixed while the incident beam and detector rotate together. The method requires a technique to fix the incident beam spot on the sample and direct the SHG signal to the detector while the optical system revolves. These requirements make this method challenging and available only recently with some variety in the setups.~\cite{14RSI_Torchinsky,15OptLett_Harter,19RSI_LuTorchinsky,20PRB_RA_SHG_Gedik}

The incidence-plane rotation method realized in the Hsieh group in 2014 was operated with a diffractive-optic based stop-start scheme.~\cite{14RSI_Torchinsky} The incidence was made oblique by diffracting the original ray with a phase mask and refocusing the offset ray on the sample surface with a reflective objective. The SHG signal was recorded while rotating the detector and the phase mask step by step. The system was renovated to solve the power-fluctuation induced low-frequency-noise issue, by using high-speed ($\sim$4~Hz) rotating phase mask based optics and position sensitive photodetector.~\cite{15OptLett_Harter} Subsequently, the Torchinsky group~\cite{19RSI_LuTorchinsky} developed reflective optics based rotating incident beam generator for the wavelength independent measurement. However, this high speed rotation design still requires a precise alignment process to make sure the high speed rotating optics is appropriately positioned.

Here, we present another design to perform the incidence-plane rotation SHG measurement. The underlying design concept is to have a compact, stable and versatile system. Our setup is based on Fresnel rhomb optics, a PMT detector, and a femtosecond laser source composed of optical fibers doped with ytterbium (Yb). The Fresnel rhomb optics is compact, wavelength independent, and misalignment-tolerant in principle. Although we applied the stop-start-type stepping-motor method, the low frequency noise is under control with the hour-long stability of the compact fiber-laser source. 

This paper is organized as follows. After the present introduction (Sec.~\ref{sec_Intro}), we describe the optics to rotate the incidence plane in Sec.~\ref{sec_RotOptics}. In Sec.~\ref{sec_ANDi}, we sketch the home-made fiber laser source and its stability. In Sec.~\ref{sec_setup}, we describe the assembly and alignment of the entire system. In Sec.~\ref{sec_demo}, we show the SHG signal data of a gold mirror and check the operation of the SHG optics (Sec.~\ref{ssec_Au}), and also present the SHG pattern of a few-layer WSe$_2$, from which we infer to the crystal orientation and degree of local effects (Sec.~\ref{ssec_WSe2}). The summary and outlook are provided in Sec.~\ref{sec_discussion}. In Appendixes A and B, we estimate the intensity of the faint SHG signal of gold, the order of which corresponds to 0.1 photon per pulse (Appendix~\ref{App_Wattage}), and subsequently describe the SHG tensor analysis at the oblique incidence (Appendix~\ref{App_tensor}).

\section{The rotary optics}
\label{sec_RotOptics}

In this section, we first sketch the essential features to rotate the incidence plane and subsequently describe the specific optics that we have adopted. 

\begin{figure}[htb]
\includegraphics{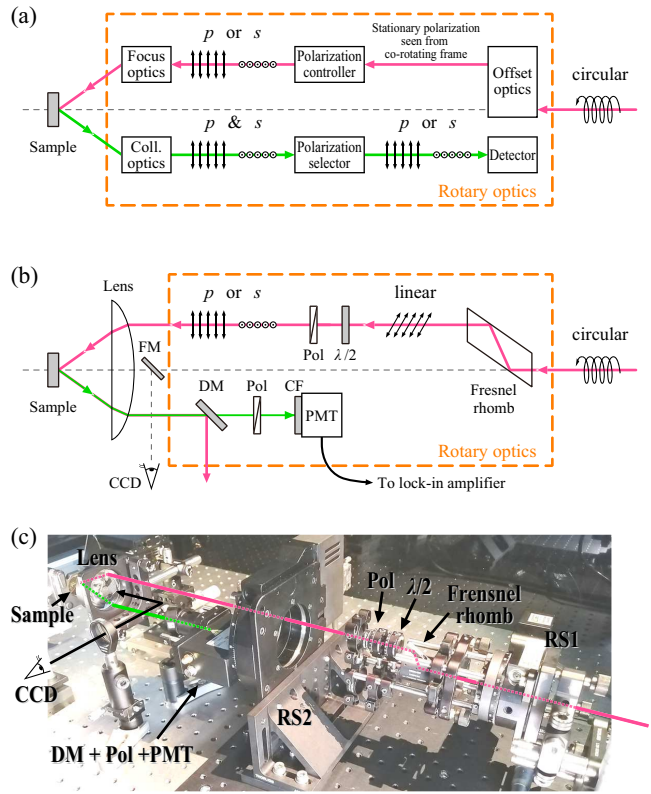}
\caption{\label{fig_RotOptics} Rotary optics. (a) Essential features in the rotary optics. (b) Optics adopted in the present setup. (c) A snapshot of the rotary optics (Multimedia view).}
\end{figure}

In the incidence-plane rotation method, there are some optical components that rotate about the axis normal to the sample surface [Fig.~\ref{fig_RotOptics}(a)]. The role of the rotary optics is to (1) rotate the incidence plane, (2) set the polarization of the incident beam, (3) focus the beam on the sample, (4) filtrate the SHG signal and its polarization and (5) direct the filtered signal to the detector. 

During the turn of the rotary optics, the direction of the incident polarization also has to follow the turn and maintain the \textit{p}- or \textit{s}-polarized incidence on the sample surface (Fig.~\ref{fig_IncRot}). To this end, it is convenient to inject the circularly polarized beam into the rotary optics along the rotation axis; see the right side of the schematic of Fig.~\ref{fig_RotOptics}(a). The convenience is due to the rotational symmetry of the circular polarization; the polarization (Jones) vector for the circular polarization is stationary seen from both the laboratory frame and the frame fixed on the rotary optics. As a result, whatever the rotation angle of the rotary optics may be, the polarization vector of the circularly-polarized injection is stationary when seen from the frame fixed to the rotary optics. 

The essential features in the rotary optics are illustrated in the dashed rectangular region in Fig.~\ref{fig_RotOptics}(a). The injected beam is first shifted from the roto-axis by the offset optics. Subsequently, the polarization is set to either \textit{p} or \textit{s} by the polarization controller. Finally, through the focus optics, the beam is directed to the sample surface at an oblique angle. The SHG signal from the surface is directed to the detector through the collimating optics. Before reaching the detector, the signal goes through the polarization selector that filtrates the \textit{p}- or \textit{s}-polarization. 

Figure~\ref{fig_RotOptics}(b) shows the optics adopted in our setup. For the sake of compactness, we used a Fresnel rhomb (FR600QM, Thorlabs) at the entrance of the rotary optics. The Fresnel rhomb offsets the injected beam for 16~mm and simultaneously transforms the circular polarization into linear over a wide range of wavelength. Note that the beam ejected from the Fresnel rhomb need not be parallel to the roto-axis, and this degree of freedom increases the tolerance of the system against alignment error. 

The subsequent polarization controller can be a half wave plate ($\lambda/2$) that rotates the linear polarization. In our setup, we also added a nanoparticle-film-type polarizer (Pol) to ensure the linear polarization. 

For the focus and collimating optics, we used one large plano-convex lens of 50~mm in diameter and a focal length of 38~mm, which sets the incidence angle to $\theta=$ 23$^{\circ}$. Instead of attaching the lens on the rotator, we fixed it on the optical table, thanks to the rotational symmetry of the lens about the lens axis. As a result of the fixing, there is a gap between the lens and the rotary optics. This gap can be utilized for the purpose to view the sample through a charge-coupled-device (CCD) camera (Sec.~\ref{sec_setup}). Moreover, this gap opens a pathway for time-resolved measurements:~\cite{14NCom_Ferro_TrSHG,18AdvMater_TSHG_MoSe} We can introduce a beam splitter in the gap and direct the second beam to the sample at the normal incidence without blocking the incident and SHG paths; here, the splitter can be held by a large transparent window fixed to the optical table. 

We adopted a PMT (PMM01, Thorlabs) for the detector. The detector is protected from the strong excitation beam by a dichroic mirror (DM) and a color filter (CF), and a Glan-Taylor-type cubic polarizer was added to select the polarization of the SHG. The beam-induced signal is modulated in $\sim$2~kHz by a mechanical chopper inserted in the injection beamline (not shown in the schematic) and the modulated current from the PMT is read with a lock-in amplifier. 

The rotary optics consists of two stepping-motor rotary stages, RS1 and RS2 (OSMS-60YAW and OSMS-120YAW, Sigma Koki) as shown in Fig.~\ref{fig_RotOptics}(c). RS1 holds a cage system on which the optical components to rotate the incident beam are attached, and RS2 is for the SHG receiving optics including the PMT detector. The two rotary stages are computor controlled and co-rotate during the data acquisition. The rotary optics including the sample stage occupied an area of 50 $\times$ 100~cm$^2$.

\section{The fiber laser source}
\label{sec_ANDi}

Optical fibers can be wound into a spool and therefore using them in a laser cavity is suited for designing a compact laser source. Besides, fiber lasers are stabe due to the all-solid nature and also cost effective owing to the economics scale of the telecommunications industry.~\cite{13NPhoton_Rev_UltrafastFiber} For the SHG measurement system, we adopted an all-normal-dispersion (ANDi) femtosecond fiber laser source.~\cite{06OptExp_ANDi} As will be shown in Sec.~\ref{sec_demo}, the pulse with the energy as large 8~nJ obtained directly at 10~MHz was strong and frequent enough to generate detectable SHG signals of gold and thin WSe$_2$ samples. 

\begin{figure}[htb]
\includegraphics{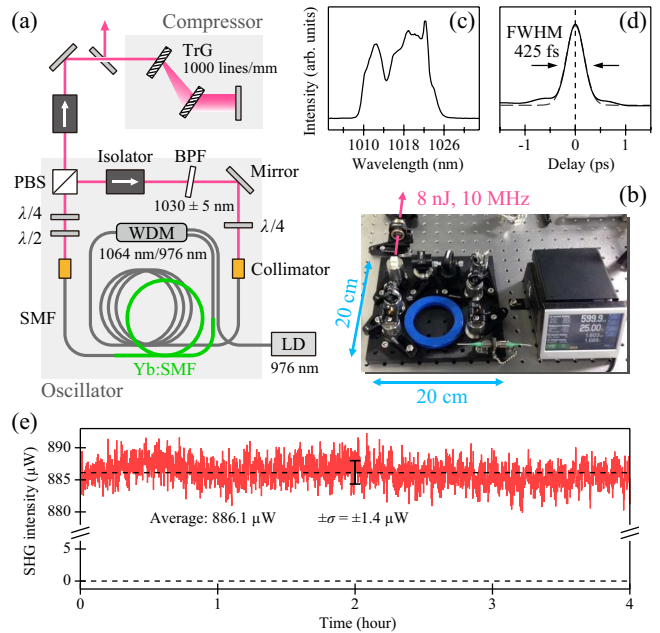}
\caption{\label{fig_ANDi} Femtosecond laser source. (a) Schematic of the oscillator and compressor. (b) Snapshot of the oscillator and LD. (c, d) Spectrum (c) and auto-correlation trace (d) of the pulse after the compressor. The trace in (d) is overlaid with a Gaussian (dashed curve) with FWHM of 425~fs. (e) Stability of the SHG of a BBO crystal. The error bar is $\pm$$\sigma$ about the average indicated by a dotted horizontal line.}
\end{figure}

Figure~\ref{fig_ANDi}(a) shows the configuration of the ANDi laser. A $\sim$20-m-long ring cavity is compacted on a board of 20 $\times$ 20cm$^2$ as shown in Fig.~\ref{fig_ANDi}(b). The Yb-doped single mode gain fiber (Yb:SMF) spliced into the single mode fiber of a core diameter of 6~$\mu$m is excited by a 976-nm wavelength laser diode (LD; BL976-SAG300, Thorlabs) through a wavelength division multiplex (WDM). In the 30-cm-long free space between two collimators, there are two quarter wave plates ($\lambda/4$), a half wave plate ($\lambda/2$) and a band pass filter (BPF) of 1030 $\pm$ 5~nm. The tilt of BPF and the azimuth angle of each of the wave plates are adjusted for the cavity to be mode locked. An isolator and a polarizing beam splitter (PBS) are also inserted in the free space; the former ensures the pulse to circulate the ring in one-way and the latter extracts a fraction of the circulating pulse from the ring cavity. The extracted laser pulse from the cavity passes another isolator that protects the oscillator from unwanted retro-reflected beam. Subsequently, the pulse passes through a pair of transmission gratings (TrG's) with 1000 grooves/mm that compresses the pulse. 

To optimize the mode-locked state of the cavity for the SHG measurement, we directed the beam after the compressor into a nonlinear crystal $\beta$-BaB$_2$O$_4$ (BBO), monitored the intensity of the SHG signal generated therefrom, and searched for the state where the signal intensity was maximized. The optimal state to our best effort was found when the 976-nm input was $\sim$400~mW, or around the maximum attained by the LD, and when the tilt of BPF was $\sim$9$^{\circ}$; the tilt can be seen in Fig.~\ref{fig_ANDi}(b). The output from the cavity was 80~mW and the repetition frequency was 10~MHz, giving 8~nJ per pulse. Figure~\ref{fig_ANDi}(c) shows the spectrum of the pulse. The spectrum was centered at 1018~nm and the bandwidth was around 15~nm. The center wavelength was shifted from the nominal value of 1030~nm for the BPF due to the $\sim$9$^{\circ}$ tilt. Figure~\ref{fig_ANDi}(d) shows the auto-correlation trace of the pulse after the compressor; overlaid on the trace is the Gaussian function whose full width at half the maximum (FWHM) was 425~fs, which corresponds to the pulse FWHM of 301~fs when the temporal profile of the pulse is assumed to be a Gaussian. Because of the rather high input of $\sim$400~mW, the mode-locked pulse was presumably chirped nonlinearly during the round trip in the cavity, and therefore, the compression of the pulse to the transform limit was difficult to be achieved; we note that the estimated limit of the auto-correlation FWHM was 203~fs for the spectrum shown in Fig.~\ref{fig_ANDi}(c). At input powers lower than 400~mW, there were mode-locked states that resulted in sub-300-fs pulses although the SHG signal intensity from BBO was weaker than the state found at 400~mW.

To show the power stability, we monitored the intensity of the SHG signal generated by the BBO crystal. The intensity monitored for 4~hours was averaged as 886.1~$\mu$W with one standard deviation $\sigma=$ 1.4~$\mu$W [Fig.~\ref{fig_ANDi}(e)]; namely, the power fluctuation was less than 0.2~\%. The low fluctuation helps to balance the vulnerability of the minute-long stop-start-type measurement to low-frequency noise.

\section{Assembly and alignment of the system}
\label{sec_setup} 

The SHG detection system consists of the rotary optics and laser source. The entire system including the sample stage can be packed in the table area of 1 $\times$ 1~m$^2$. 

There are four essential points when assembling and aligning the system: (1) The beam injected into the rotary optics has to be highly circularly polarized. (2) The injection beam has to be co-axial to the roto-axis of the rotary optics. (3) The incident beam has to be directed to the point on the sample surface where the roto-axis intersects. (4) The sample surface has to be perpendicular to the roto-axis. Once the requirements (2) to (4) are met, the SHG signal will have a fixed trajectory when seen from the frame on the rotary optics. In our setup, the rotary optics consists of two stepping-motor rotary stages and one plano-convex lens, as explained in Sec.~\ref{sec_RotOptics}. Therefore, aligning these three separate parts to form a firm roto-axis is the prerequisite to the four criteria mentioned above. 

Here, we describe some of the practical procedures to assemble and align the SHG measurement system. 

First, we retract the plano-convex lens facing the sample stage, remove the optics on the cage held by the stepping motor rotary stage RS1, and also detach the optical components including the PMT detector from RS2. At this point, we check the co-axial alignment of RS1 and RS2. Then, by using some irises, we let the injection beam path through the roto-centers of the two stepping motors, RS1 and RS2. This beam path becomes the roto-axis. For the purpose of alignment, we also set another laser beam that counter-propagates along the roto-axis. 

Next, we attach a gold mirror on the sample stage and adjust its orientation to retro-reflect the injected beam so that the mirror surface becomes perpendicular to the roto-axis. The gold mirror will serve as a reference sample for checking the alignment of the SHG setup; see Sec.~\ref{sec_demo}. 

Then, we translate the plano-convex lens into the optical path. The tilt and position of the lens have to be adjusted so that the lens axis becomes co-axial to the roto-axis. To adjust the tilt, we retract the gold mirror and illuminate the planer side of the lens with the counter-propagating beam that has been aligned co-axial to the prinicipal axis. We adjust the tilt until the faint reflection from the planer surface of the lens is retro-reflected. To adjust the position, we re-insert the gold mirror and adjust the position of the lens until the reflection from the gold mirror matches the injection beam path. 

Next, we attach the incidence beam optics, namely, the Fresnel rhomb, half wave plate, and thin-film polarizer, on the cage held by the rotary stage RS1. We also attach a small power meter on the cage and monitor the power of the offset and polarization-controlled beam while RS1 is in motion. We adjust the quarter wave plate in the injection beamline [located outside the area seen in Fig.~\ref{fig_RotOptics}(c)] until the variation of the monitored power during the rotation of RS1 becomes less than $\pm$1~\%. This procedure is to ensure the circular polarization of the injection beam. 

After setting the circular polarization, we remove the small power meter from the cage and direct the beam on the sample surface at an oblique angle. There is a flip mirror (FM) in the gap between the lens and the rotary optics; the FM opens a path to view the sample through the CCD camera [Figs.~\ref{fig_RotOptics}(b) and \ref{fig_RotOptics}(c)]. Note that the incidence beam spot on the sample can be viewed with the CCD because of its sensitivity in the infra-reds. The beam spot generally draws a circle as the incidence plane is rotated. We shift the sample along the roto-axis until the beam spot stays still on a point. This point is identified as the intersection of the roto-axis and the sample surface. 

Finally, we flip out the FM and attach the SHG receiving optics on the rotary stage RS2.

\section{Demonstration of the SHG measurement}
\label{sec_demo}

\subsection{The SHG of a gold mirror as a reference}
\label{ssec_Au}

When the SHG measurement system is perfectly aligned, the SHG signal from the gold mirror would be constant over the 360$^{\circ}$ rotation of the incidence plane because the isotropic polycrystalline mirror surface effectively creates rotational isotropy characterized by the point group symmetry of ${\rm C}_{\infty {\rm v}}$. The degree of isotropy of the SHG signal of the gold mirror, thus, can serve as the measure of the alignment accuracy. 

\begin{figure}[htb]
\includegraphics{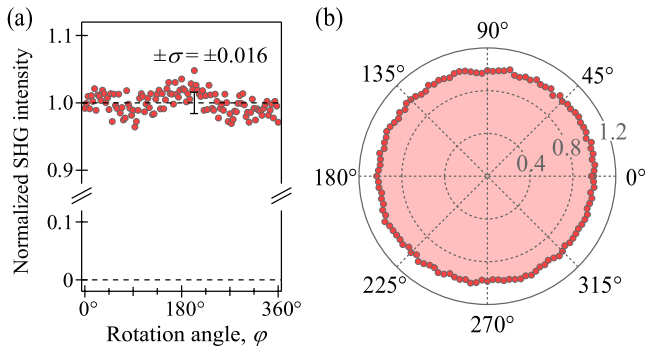}
\caption{\label{fig_Au} SHG of a gold mirror. (a) Normalized SHG intensity as a function of the rotation angle $\varphi$. The error bar indicates one standard deviation $\sigma$ = 0.016 of the 121 data points. (b) The polar plot of (a).}
\end{figure}

\begin{figure*}[htb]
\includegraphics{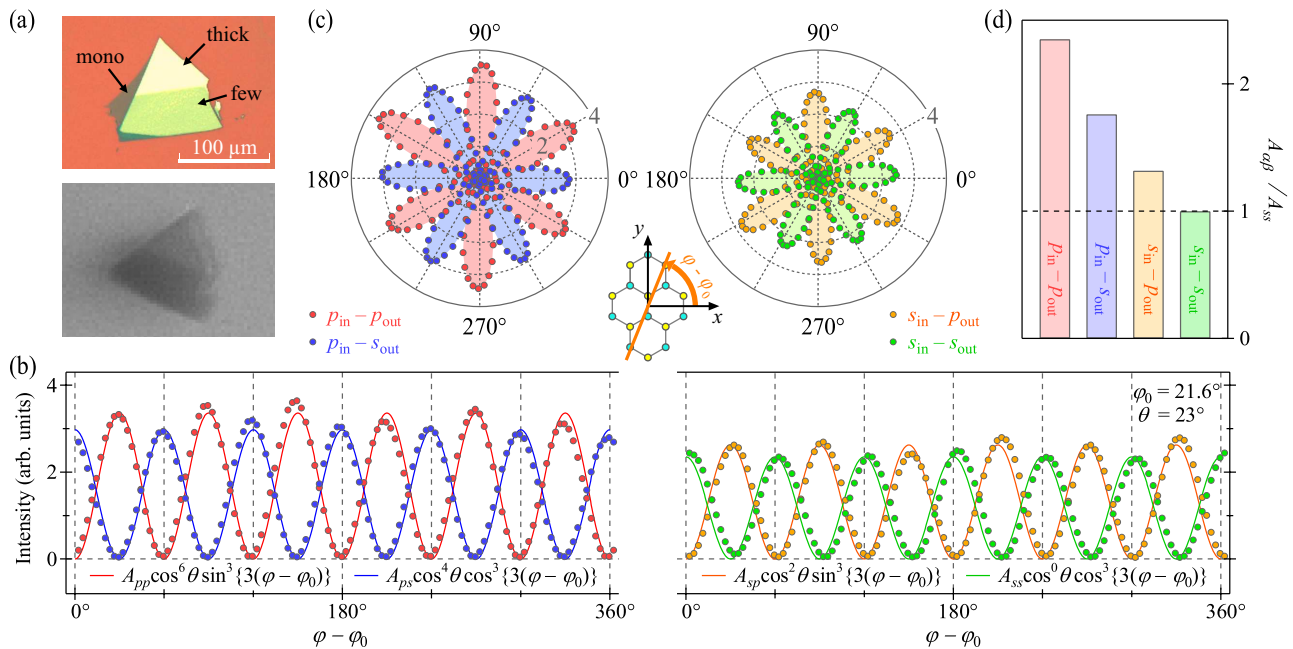}
\caption{\label{fig_WSe2} SHG of a few-layers WSe$_2$ film. (a) Sample seen through an optical microscope (upper image) and the identical sample on the sample stage seen through the CCD camera (lower image). Mono-, few-, and thick-layer areas on the sample are indicated by arrows. (b) SHG signal-intensity profiles for the four polarization configurations as functions of the rotation angle. Theoretical curves are overlaid on the data points. (c) Polar plot of (b). (d) Amplitude of the SHG pattern normalized to $A_{ss}$.The dashed horizontal line is the expected level when there is no local-field effects. The middle inset shows the hexagonal crystal surface. The $y$ axis is set along the arm-chair direction, which is in the mirror plane; here, the setting of the Cartsian coordinate system follows the indexing adopted for $\chi^{(2)}_{ijk}$ tabulated in the textbook.~\cite{Boyd} }
\end{figure*}

The measurement was performed at room temperature. The beam spot diameter on the sample surface was 20~$\mu$m as estimated from the beam seen through the CCD camera. The incident beam was \textit{p}-polarized with the incidence angle of 23$^{\circ}$; its wattage was 30~mW and the spot area on the sample was 300~$\mu$m$^2$ so that the fluence was 1~mJ/cm$^2$; the \textit{p}-polarized SHG was detected. The acquisition time for the 121 data points with the azimuth-angular step of 3$^{\circ}$ was $\sim$6~min.

Figure~\ref{fig_Au}(a) shows the SHG signal intensity as a function of the rotation angle $\varphi$ of the incidence plane. First of all, the SHG intensity under the setup was intense enough to be detected by the PMT read with the lock-in amplifier. The estimated wattage of the SHG was 0.5~pW, which is the order of 0.1 photon per pulse; for the detection scheme of the small wattage, see Appendix~\ref{App_Wattage}. Second, the intensity profile after the alignment was fairly isotropic in the radial plot [Fig.~\ref{fig_Au}(b)]. The intensity is constant with one standard deviation $\sigma$ less than 2~\% of the average value. The degree of the isotropy is the measure of the integrity of the setup, and we routinely check the isotropy before measuring the samples.

\subsection{The SHG of a few-layer WSe$_2$}
\label{ssec_WSe2}

As a demonstration, we measured the SHG profile of a few-layer WSe$_2$ sample. The atomically thin transition metal dichalcogenides including WSe$_2$ with ${\rm D}_{3{\rm h}}$ symmetry are known to exhibit strong SHG emissions.~\cite{13PRB_SHG_MoS2_Kumar,13PRB_SHG_MoS2_Malard,13NanoLett_SHG_MoS2_Heinz,14SciRep_SHG_WS2,15TwoDMater_SHG_WSe2,18SciRep_SHG_WSe2} 

The flakes of WSe$_2$ were mechanically exfoliated on polydimethylsiloxane (PDMS) and subsequently transferred to a 300-nm SiO$_2$/Si substrate. The upper image of Fig.~\ref{fig_WSe2}(a) shows one of the transferred flakes seen through an optical microscope. The area of $\sim$100 $\times$ 50~$\mu$m$^2$ was identified to have a few mono-layers; see the arrows in the upper image of Fig.~\ref{fig_WSe2}(a) indicating the mono-, few- and thick-layer areas. After the characterization by using the microscope, the sample was mounted on the sample stage. The lower image of Fig.~\ref{fig_WSe2}(a) is the mounted sample seen via the CCD camera. We adjusted the sample position so that the incidence beam spot stayed on the few-layer area while the rotary stage was in motion. The SHG measurement was performed at room temperature, and the incident fluence was 1~mJ/cm$^2$ as in the gold mirror measurement. The incidence polarization was set to \textit{p} (\textit{p}$_{\rm in}$) and both the \textit{p}- and \textit{s}-polarized SHG (\textit{p}$_{\rm out}$ and \textit{s}$_{\rm out}$) were recorded.

Figure~\ref{fig_WSe2}(b) shows the SHG profile of the few-layer WSe$_2$ sample, and Fig.~\ref{fig_WSe2}(c) is its polar plot. The six-fold SHG pattern is observed for all the polarization configurations. First, the symmetric pattern indicates that the beam spot stayed on the $\sim$100 $\times$ 50~$\mu$m$^2$ area on the sample surface during the turn of the rotary optics. 

The standard analysis of the SHG pattern relies on the symmetry of the second-order nonlinear optical polarizability, $\chi_{ijk}^{(2)}$. When the crystal is symmetric with the $\rm{D}_{3{\rm h}}$ point-group operations, there are only four non-zero components of $\chi_{ijk}^{(2)}$ which are related to each other as~\cite{Boyd} 
\begin{equation}\label{eq_chi}
\chi^{(2)}_{yyy} = -\chi^{(2)}_{yxx} = -\chi^{(2)}_{xxy} = -\chi^{(2)}_{xyx} \equiv \chi.
\end{equation}
Because $\chi^{(2)}_{yyy}$ is finite,~\cite{16RevMex_Chi2} $yz$ is the mirror plane in the Cartesian indexing adopted in Eq.~(\ref{eq_chi}), and thus, $y$ is along the armchair direction while $x$ is along the zig-zag direction of the hexagonal WSe$_2$ crystal; see the middle inset of Fig.~\ref{fig_WSe2}. We also note that the $z$ component of the polarization is not induced because $\chi^{(2)}_{ijk}$ vanishes for $i=z$. The SHG signal intensity $I_{\alpha\beta}$ for the $\alpha_{\rm in}$-$\beta_{\rm out}$ configuration ($\alpha,\beta=$ \textit{p} or \textit{s}) can be calculated as functions of the rotation angle $\varphi-\varphi_0$ and incidence angle $\theta$ as (see Fig.~\ref{fig_IncRot} and Appendix~\ref{App_tensor})
\begin{eqnarray}
I_{\textit pp} &=& A_{\textit pp} \cos^6\theta \sin^2\{3(\varphi-\varphi_0)\},\label{eq_I1}\\
I_{\textit ps} &=& A_{\textit ps} \cos^4\theta \cos^2\{3(\varphi-\varphi_0)\},\label{eq_I2}\\
I_{\textit sp} &=& A_{\textit sp} \cos^2\theta \sin^2\{3(\varphi-\varphi_0)\},\label{eq_I3}\\
I_{\textit ss} &=& A_{\textit ss} \cos^0\theta \cos^2\{3(\varphi-\varphi_0)\}.\label{eq_I4}
\end{eqnarray}
Here, $A_{\alpha\beta}$ is a proportionality constant which includes the effects of local fields~\cite{14SciRep_SHG_WS2,16JOptSocAmB_SHG_ZincBlend} and that can, in principle, depend on the configuration of the light (Appendix~\ref{App_tensor}).

In Fig.~\ref{fig_WSe2}(b), we overlay the sinusoidal curves of Eqs.~(\ref{eq_I1}) to (\ref{eq_I4}) with $\theta=$ 23$^{\circ}$ and $\varphi_0=$ 21.6$^{\circ}$. The curves nicely reproduce the angular pattern. The relative amplitudes $A_{\alpha\beta}/A_\textit{ss}$ derived from the fit are plotted in Fig.~\ref{fig_WSe2}(d). More than a factor of 2 difference was observed among the four amplitudes. The amplitude can depend on the polarizatoin through the Fresnel and local-field effects (Appendix~\ref{App_tensor}). Because the Fresnel effect, or the difference in the reflection and transmission between $p$- and $s$-polarizations, is negligible at $\theta$ as small as 23$^{\circ}$, we attribute the strong dependence to the local-field effects. In other words, the relative amplitudes of the SHG signal among the four polarization configurations fingerprint whether the degree of the local-field effects is substantial or not, the information of which is uniquely derived using the off-normal incidence-plane rotation method.

\section{Summary and outlook}
\label{sec_discussion}
In summary, we have constructed a compact and stable incidence-plane-rotating second-harmonics detector based on Fresnel-rhomb optics, a PMT detector, and a home-made Yb:fiber laser source. We described the design principle and schematics (Sec.~\ref{sec_RotOptics} and \ref{sec_ANDi}), alignment procedure (Sec.~\ref{sec_setup}), and experimental realization (Sec.~\ref{sec_demo}) of the setup. The entire system fits into the optical table area of 1 $\times$ 1~m$^2$ and can produce harmonic signal with the intensity deviation less than $\pm$0.2~\% in 4~hours owing to the stability of the laser source (Fig.~\ref{fig_ANDi}). The data acquisition for a gold reference sample took $\sim$300~s for 3$^{\circ}$-step measurement and provided $<$$\pm$2~\% reliable data for the 0.5~pW output, which corresponds to $\sim$0.1~photon per pulse at the 10-MHz repetition (Fig.~\ref{fig_Au} and Appendix~\ref{App_Wattage}). The SHG data of an exfoliated WSe$_2$ film were also presented; the six-fold pattern and the relative strength for the four polarization configurations provided us with the information of the crystal orientation and degree of local effects in the atomically thin sample (Fig.~\ref{fig_WSe2} and Appendix~\ref{App_tensor}). 

Because the sample is fixed in the laboratory frame, there are more degrees of freedom to subject the sample to various conditions than the case where the sample rotates. In our setup, the working distance between the plano-convex objective lens and the sample is around 38~mm, and therefore, cryostats can readily be installed to cool down the sample. In addition, because we have set the objective lens fixed instead of rotating, there is also a gap between the lens and the rotary optics; this gap can be utilized for the pump-probe measurement, as described in Sec.~\ref{sec_setup}. Thus, our incidence-plane rotation detection design can be extended to investigating the samples under various conditions such as cryogenic, strained, ultrafast non-equilibrium, and high magnetic fields.

\begin{acknowledgments}
This work was conducted under the ISSP-CCES Collaborative Program and was supported by the Institute for Basic Science in Republic of Korea (Grant Numbers IBS-R009-Y2 and IBS-R009-G2) and by JSPS KAKENHI (Grant Numbers 17K18749, 18H01148, 19K22140 and 19KK0350). 
\end{acknowledgments}

\appendix

\section{Measuring the faint SHG wattage}
\label{App_Wattage}

The SHG wattage of the gold mirror $I_{2\omega} =$ 0.5~pW is at the scale of 0.1 photon per pulse: one SHG photon of 509~nm emitted at 10~MHz corresponds to 
3.9~pW. The faint wattage was measured by utilizing an artificial SHG emission of a BBO crystal as described below. 

The PMT sensitivity (signal voltage per a fixed photon count/s) is controlled by the gain control voltage: When we increase it, the gain becomes approximately 10 times higher for every 0.10~V. We first tune the gain to measure the SHG of the gold mirror; namely, we set the value large enough to detect faint SHG emission and small enough to prevent any unwanted electronic signal saturation. The lock-in amplifier modulated signal was 0.45~mV under the gain control voltage 0.80~V. We then install the PMT (with same color filter to block the fundamental beam) in front of the fundamental pulse laser beam and BBO crystal to make the artificial SHG emission. The beam need not be focused since what we want is the SHG emission of $\sim$$\mu$W at most. We control the intensity of the fundamental beam with strong continuous revolving neutral-density (ND) filter of optical density 4 until the lock-in amplifier modulated signal becomes the same value of the gold film SHG, and measure the fundamental beam intensity with a conventional power meter. We then take out the ND filter to obtain the maximum value of fundamental beam intensity with a power meter, and its resulting SHG wattage with the power meter covered with the same color filter. Note that the SHG induced by the maximum power is strong enough to be detected by the conventional power meter, and even visible as a green spot on a paper by naked eyes. 

Here, we use the fact that the SHG wattage is proportional to the square of fundamental beam intensity: $I_{2\omega}/I^{\rm max}_{2\omega} = (I_{\omega}/I^{\rm max}_{\omega})^2$. The symbols and measured values are the followings: the faint fundamental wattage $I_{\omega}$ = 64~$\mu$W; the maximum fundamental beam wattage $I^{\rm max}_{\omega}$ = 40~mW; and the maximum-power-induced SHG wattage $I^{\rm max}_{2\omega}$ = 0.25~$\mu$W. Thus, we obtain $I_{2\omega}$ = 0.5~pW, which is the faint SHG wattage of the gold mirror. 

\section{SHG tensor analysis at an oblique incidence}
\label{App_tensor}
The standard analysis of the SHG pattern of non-centrosymmetric crystals~\cite{Boyd} relies on the symmetry arguments of the second-order nonlinear optical polarizability, $\chi_{ijk}^{(2)}$. Here, we calculate the SHG signal intensity as a function of the rotation angle $\varphi-\varphi_0$ when the incidence angle $\theta$ is finite. We hereafter set $\varphi_0=$ 0 without the loss of generality and also abbreviate sine and cosine functions as $\sin\theta\to\rm{s}\theta$ and $\cos\theta\to\rm{c}\theta$, respectively. 

The polarized incidence fields $\mathbf{E}^\textit{p}(\omega)$ and $\mathbf{E}^\textit{s}(\omega)$ can be described using the unit vectors $\mathbf{\bar{e}}_{\omega}^\textit{p}$ and $\mathbf{\bar{e}}_{\omega}^\textit{s}$ directed along the polarization (see Fig.~\ref{fig_IncRot}) and further be expanded with the Cartesian coordinate basis $(\mathbf{\bar{e}}_\textit{x},\mathbf{\bar{e}}_\textit{y},\mathbf{\bar{e}}_\textit{z})$ as
\begin{equation}\label{eq_inc}
\mathbf{E}^{\alpha}(\omega)=E_\omega\mathbf{\bar{e}}_{\omega}^{\alpha}=\sum_{i=x,y,z}E_{\alpha}^i\mathbf{\bar{e}}_i. 
\end{equation}
Here, $\alpha$ (= \textit{p} or \textit{s}) indicates the polarization, $E_\omega$ is the signed amplitude and $E_{\alpha}^i$ is the Cartesian coordinate. The unit polarization vectors can be expanded with the Cartesian coordinate basis as follows (see Fig.~\ref{fig_IncRot}):
\begin{align}
\qquad\qquad
\mathbf{\bar{e}}_{\omega}^\textit{p}&	=&	-\rm{c}\theta\rm{c\varphi}&	\,\mathbf{\bar{e}}_\textit{x}&	-\rm{c}\theta\rm{s}\varphi&	\,\mathbf{\bar{e}}_\textit{y}&	+\rm{s}\theta& \,\mathbf{\bar{e}}_\textit{z},\label{eq_Inc1}\\
\mathbf{\bar{e}}_{\omega}^\textit{s}&	=&	\rm{s}\varphi&				\,\mathbf{\bar{e}}_\textit{x}&	-\rm{c}\varphi&				\,\mathbf{\bar{e}}_\textit{y}.& & \label{eq_Inc2}
\end{align}
The components on the right-hand side of Eqs.~(\ref{eq_Inc1}) and (\ref{eq_Inc2}) are the direction cosines $E_\textit{p}^{i}/E_\omega$ and $E_\textit{s}^i/E_\omega$ of the incident fields, respectively; see Eq.~(\ref{eq_inc}). Thus, 
\begin{eqnarray}
\label{eq_dcos}
\begin{pmatrix}E_\textit{p}^x \\ E_\textit{p}^y \\ E_\textit{p}^z\end{pmatrix}
= E_\omega
\begin{pmatrix}\text{--}\rm{c}\theta\rm{c}\varphi\\\text{--}\rm{c}\theta\rm{s}\varphi\\ \rm{s}\theta\end{pmatrix}, 
\begin{pmatrix} E_\textit{s}^x \\ E_\textit{s}^y \\ E_\textit{s}^z\end{pmatrix}
= E_\omega
\begin{pmatrix}\rm{s}\varphi\\\text{--}\rm{c}\varphi\\0\end{pmatrix}.
\end{eqnarray}

The incident field $\mathbf{E}^{\alpha}(\omega)$ induces the polarization $\mathbf{P}$ of the medium in the substrate. The 2$\omega$ component of the induced polarization through the second-order nonlinear optical process can be described as 
\begin{equation}\label{eq_chi2}
\mathbf{P}_{\alpha}^{(2)}(2\omega)=\sum_{i}P_{\alpha,i}^{(2)}\mathbf{\bar{e}}_\textit{i}=\sum_{ijk}\varepsilon_0\chi^{(2)}_{ijk}E_\alpha^jE_\alpha^k\mathbf{\bar{e}}_\textit{i},
\end{equation}
where $\varepsilon_0$ is the vacuum permittivity. 

The intensity of the SHG signal emitted from the induced polarization of the medium can be described as
\begin{equation}\label{eq_SHint}
I_{\alpha\beta}=a_{\alpha\beta}|\mathbf{\bar{e}}_{2\omega}^\beta\cdot\mathbf{P}_{\alpha}^{(2)}(2\omega)|^2. 
\end{equation}
Here, $\mathbf{\bar{e}}_{2\omega}^{\beta}$ ($\beta$ = \textit{p} or \textit{s}) is the unit polarization vector of the outgoing SHG field in the reflection geometry and can be described as follows (see Fig.~\ref{fig_IncRot}): 
\begin{align}
\qquad\qquad
\mathbf{\bar{e}}_{2\omega}^\textit{p}&	=&	\rm{c}\theta\rm{c}\varphi&	\,\mathbf{\bar{e}}_\textit{x}&	+\rm{c}\theta\rm{s}\varphi&	\,\mathbf{\bar{e}}_\textit{y}&	+\rm{s}\theta&	\,\mathbf{\bar{e}}_\textit{z},\label{eq_Inc3}\\
\mathbf{\bar{e}}_{2\omega}^\textit{s}&	=&	\rm{s}\varphi&				\,\mathbf{\bar{e}}_\textit{x}&	-\rm{c}\varphi&				\,\mathbf{\bar{e}}_\textit{y}.& &\label{eq_Inc4}
\end{align}
Meanwhile, $a_{\alpha\beta}$ is the proportionality constant which includes the effects of local fields as well as the Fresnel correction~\cite{14SciRep_SHG_WS2,16JOptSocAmB_SHG_ZincBlend} and that can, in principle, depend on the polarization of the fields; thus, $a$ is indexed with $\alpha$ and $\beta$. By inserting Eqs.~(\ref{eq_dcos}), (\ref{eq_chi2}), (\ref{eq_Inc3}) and (\ref{eq_Inc4}) into Eq.~(\ref{eq_SHint}), the SHG signal intensity $I_{\alpha\beta}$ can be obtained as a function of $\theta$, $\varphi$ and $|E_\omega|$. 

When the crystal is symmetric with respect to the ${\rm D}_{3{\rm h}}$ point-group operations, the elements of $\chi^{(2)}_{ijk}$ obey Eq.~(\ref{eq_chi}) and the coordinates of $\mathbf{P}_{\alpha}^{(2)}(2\omega)$ are simplified as 
\begin{eqnarray}
\begin{pmatrix}P_{\alpha,x}^{(2)} \\ P_{\alpha,y}^{(2)} \\ P_{\alpha,z}^{(2)} \end{pmatrix}\,=\,\varepsilon_0\chi\begin{pmatrix}\text{--}2E_\alpha^xE_\alpha^y\\(E_\alpha^y)^2\,\text{--}\,(E_\alpha^x)^2\\ 0\end{pmatrix}\nonumber. 
\end{eqnarray}
Explicitly, 
\begin{eqnarray}
P^{(2)}_{\textit{p},x} =&\,\varepsilon_0\chi|E_\omega|^2&\times\{\text{--}2({\rm c}\theta)^2{\rm c}\varphi{\rm s}\varphi\},\label{eq_P1}\\
P^{(2)}_{\textit{s},x} =&\,\varepsilon_0\chi|E_\omega|^2&\times\,2\,{\rm s}\varphi{\rm c}\varphi,\label{eq_P2}\\
P^{(2)}_{\textit{p},y} =&\,\varepsilon_0\chi|E_\omega|^2&\times\{({\rm c}\theta)^2({\rm s}\varphi)^2-({\rm c}\theta)^2({\rm c}\varphi)^2\},\label{eq_P3}\\
P^{(2)}_{\textit{s},y} =&\,\varepsilon_0\chi|E_\omega|^2&\times\{({\rm c}\varphi)^2-({\rm s}\varphi)^2\},\label{eq_P4}\\
P^{(2)}_{\textit{p},z} =&\,P^{(2)}_{\textit{s},z}& =\,0.\label{eq_P5} 
\end{eqnarray}
By inserting Eqs.~(\ref{eq_Inc3}) to (\ref{eq_P5}) into Eq.~(\ref{eq_SHint}), the angular dependence of the SHG signal intensity Eqs.~(\ref{eq_I1}) to (\ref{eq_I4}) in the main text is obtained, where $A_{\alpha\beta}=a_{\alpha\beta}\varepsilon_0^2\chi^2|E_\omega|^4$.

\section*{Data availability}
The data that support the findings of this study are available from the corresponding author upon reasonable request.

\section*{References}


\begin{thebibliography}{37}%
\makeatletter
\providecommand \@ifxundefined [1]{%
 \@ifx{#1\undefined}
}%
\providecommand \@ifnum [1]{%
 \ifnum #1\expandafter \@firstoftwo
 \else \expandafter \@secondoftwo
 \fi
}%
\providecommand \@ifx [1]{%
 \ifx #1\expandafter \@firstoftwo
 \else \expandafter \@secondoftwo
 \fi
}%
\providecommand \natexlab [1]{#1}%
\providecommand \enquote  [1]{``#1''}%
\providecommand \bibnamefont  [1]{#1}%
\providecommand \bibfnamefont [1]{#1}%
\providecommand \citenamefont [1]{#1}%
\providecommand \href@noop [0]{\@secondoftwo}%
\providecommand \href [0]{\begingroup \@sanitize@url \@href}%
\providecommand \@href[1]{\@@startlink{#1}\@@href}%
\providecommand \@@href[1]{\endgroup#1\@@endlink}%
\providecommand \@sanitize@url [0]{\catcode `\\12\catcode `\$12\catcode
  `\&12\catcode `\#12\catcode `\^12\catcode `\_12\catcode `\%12\relax}%
\providecommand \@@startlink[1]{}%
\providecommand \@@endlink[0]{}%
\providecommand \url  [0]{\begingroup\@sanitize@url \@url }%
\providecommand \@url [1]{\endgroup\@href {#1}{\urlprefix }}%
\providecommand \urlprefix  [0]{URL }%
\providecommand \Eprint [0]{\href }%
\providecommand \doibase [0]{http://dx.doi.org/}%
\providecommand \selectlanguage [0]{\@gobble}%
\providecommand \bibinfo  [0]{\@secondoftwo}%
\providecommand \bibfield  [0]{\@secondoftwo}%
\providecommand \translation [1]{[#1]}%
\providecommand \BibitemOpen [0]{}%
\providecommand \bibitemStop [0]{}%
\providecommand \bibitemNoStop [0]{.\EOS\space}%
\providecommand \EOS [0]{\spacefactor3000\relax}%
\providecommand \BibitemShut  [1]{\csname bibitem#1\endcsname}%
\let\auto@bib@innerbib\@empty
\bibitem [{\citenamefont {Franken}\ \emph {et~al.}(1961)\citenamefont
  {Franken}, \citenamefont {Hill}, \citenamefont {Peters},\ and\ \citenamefont
  {Weinreich}}]{61PRL_Franken}%
  \BibitemOpen
  \bibfield  {author} {\bibinfo {author} {\bibfnamefont {P.~A.}\ \bibnamefont
  {Franken}}, \bibinfo {author} {\bibfnamefont {A.~E.}\ \bibnamefont {Hill}},
  \bibinfo {author} {\bibfnamefont {C.~W.}\ \bibnamefont {Peters}}, \ and\
  \bibinfo {author} {\bibfnamefont {G.}~\bibnamefont {Weinreich}},\ }\bibfield
  {title} {\enquote {\bibinfo {title} {{Generation of optical harmonics}},}\
  }\href@noop {} {\bibfield  {journal} {\bibinfo  {journal} {Phys. Rev. Lett.}\
  }\textbf {\bibinfo {volume} {7}},\ \bibinfo {pages} {118--119} (\bibinfo
  {year} {1961})}\BibitemShut {NoStop}%
\bibitem [{\citenamefont {Boyd}(2008)}]{Boyd}%
  \BibitemOpen
  \bibfield  {author} {\bibinfo {author} {\bibfnamefont {R.~W.}\ \bibnamefont
  {Boyd}},\ }\href@noop {} {\emph {\bibinfo {title} {{Nonlinear optics}}}},\
  \bibinfo {edition} {3rd}\ ed.\ (\bibinfo  {publisher} {Achademic Press},\
  \bibinfo {year} {2008})\BibitemShut {NoStop}%
\bibitem [{\citenamefont {Yin}\ \emph {et~al.}(2014)\citenamefont {Yin},
  \citenamefont {Ye}, \citenamefont {Chenet}, \citenamefont {Ye}, \citenamefont
  {O’Brien}, \citenamefont {Hone},\ and\ \citenamefont
  {Zhang}}]{14Sci_MoS2_Edge}%
  \BibitemOpen
  \bibfield  {author} {\bibinfo {author} {\bibfnamefont {X.}~\bibnamefont
  {Yin}}, \bibinfo {author} {\bibfnamefont {Z.}~\bibnamefont {Ye}}, \bibinfo
  {author} {\bibfnamefont {D.~A.}\ \bibnamefont {Chenet}}, \bibinfo {author}
  {\bibfnamefont {Y.}~\bibnamefont {Ye}}, \bibinfo {author} {\bibfnamefont
  {K.}~\bibnamefont {O’Brien}}, \bibinfo {author} {\bibfnamefont {J.~C.}\
  \bibnamefont {Hone}}, \ and\ \bibinfo {author} {\bibfnamefont
  {X.}~\bibnamefont {Zhang}},\ }\bibfield  {title} {\enquote {\bibinfo {title}
  {{Edge nonlinear optics on a MoS$_2$ atomic monolayer}},}\ }\href@noop {}
  {\bibfield  {journal} {\bibinfo  {journal} {Science}\ }\textbf {\bibinfo
  {volume} {344}},\ \bibinfo {pages} {488--490} (\bibinfo {year}
  {2014})}\BibitemShut {NoStop}%
\bibitem [{\citenamefont {Shen}(1989)}]{89YRShen}%
  \BibitemOpen
  \bibfield  {author} {\bibinfo {author} {\bibfnamefont {Y.~R.}\ \bibnamefont
  {Shen}},\ }\bibfield  {title} {\enquote {\bibinfo {title} {{Optical second
  harmonic gereration at interfaces}},}\ }\href@noop {} {\bibfield  {journal}
  {\bibinfo  {journal} {Annu. Rev. Phys. Chem.}\ }\textbf {\bibinfo {volume}
  {40}},\ \bibinfo {pages} {327--350} (\bibinfo {year} {1989})}\BibitemShut
  {NoStop}%
\bibitem [{\citenamefont {Heinz}(1991)}]{91Heinz}%
  \BibitemOpen
  \bibfield  {author} {\bibinfo {author} {\bibfnamefont {T.~F.}\ \bibnamefont
  {Heinz}},\ }\bibfield  {title} {\enquote {\bibinfo {title} {{Second-order
  nonlinear optical effects at surfaces and interfaces}},}\ }\href@noop {}
  {\bibfield  {journal} {\bibinfo  {journal} {{\rm in} {\it Nonlinear Surface
  Electromagnetic Phenomena}, eds.\ H. E. Ponath \& G. I. Stegeman}\ ,\
  \bibinfo {pages} {353--416}} (\bibinfo {year} {North Holland, Amsterdam,
  1991})}\BibitemShut {NoStop}%
\bibitem [{\citenamefont {Hsieh}\ \emph
  {et~al.}(2011{\natexlab{a}})\citenamefont {Hsieh}, \citenamefont {McIver},
  \citenamefont {Torchinsky}, \citenamefont {Gardner}, \citenamefont {Lee},\
  and\ \citenamefont {Gedik}}]{11PRL_Hsieh_BiSe_Aging}%
  \BibitemOpen
  \bibfield  {author} {\bibinfo {author} {\bibfnamefont {D.}~\bibnamefont
  {Hsieh}}, \bibinfo {author} {\bibfnamefont {J.~W.}\ \bibnamefont {McIver}},
  \bibinfo {author} {\bibfnamefont {D.~H.}\ \bibnamefont {Torchinsky}},
  \bibinfo {author} {\bibfnamefont {D.~R.}\ \bibnamefont {Gardner}}, \bibinfo
  {author} {\bibfnamefont {Y.~S.}\ \bibnamefont {Lee}}, \ and\ \bibinfo
  {author} {\bibfnamefont {N.}~\bibnamefont {Gedik}},\ }\bibfield  {title}
  {\enquote {\bibinfo {title} {{Nonlinear optical probe of tunable surface
  electrons on a topological insulator}},}\ }\href@noop {} {\bibfield
  {journal} {\bibinfo  {journal} {Phys. Rev. Lett.}\ }\textbf {\bibinfo
  {volume} {106}},\ \bibinfo {pages} {057401} (\bibinfo {year}
  {2011}{\natexlab{a}})}\BibitemShut {NoStop}%
\bibitem [{\citenamefont {Hsieh}\ \emph
  {et~al.}(2011{\natexlab{b}})\citenamefont {Hsieh}, \citenamefont {Mahmood},
  \citenamefont {McIver}, \citenamefont {Gardner}, \citenamefont {Lee},\ and\
  \citenamefont {Gedik}}]{11PRL_Hsieh_BiSe_Dynamics}%
  \BibitemOpen
  \bibfield  {author} {\bibinfo {author} {\bibfnamefont {D.}~\bibnamefont
  {Hsieh}}, \bibinfo {author} {\bibfnamefont {F.}~\bibnamefont {Mahmood}},
  \bibinfo {author} {\bibfnamefont {J.~W.}\ \bibnamefont {McIver}}, \bibinfo
  {author} {\bibfnamefont {D.~R.}\ \bibnamefont {Gardner}}, \bibinfo {author}
  {\bibfnamefont {Y.~S.}\ \bibnamefont {Lee}}, \ and\ \bibinfo {author}
  {\bibfnamefont {N.}~\bibnamefont {Gedik}},\ }\bibfield  {title} {\enquote
  {\bibinfo {title} {{Selective probing of photoinduced charge and spin
  dynamics in the bulk and surface of a topological insulator}},}\ }\href@noop
  {} {\bibfield  {journal} {\bibinfo  {journal} {Phys. Rev. Lett.}\ }\textbf
  {\bibinfo {volume} {107}},\ \bibinfo {pages} {077401} (\bibinfo {year}
  {2011}{\natexlab{b}})}\BibitemShut {NoStop}%
\bibitem [{\citenamefont {Benalcazar}, \citenamefont {Bernevig},\ and\
  \citenamefont {Hughes}(2017)}]{17Sci_HOTI_Theory}%
  \BibitemOpen
  \bibfield  {author} {\bibinfo {author} {\bibfnamefont {W.~A.}\ \bibnamefont
  {Benalcazar}}, \bibinfo {author} {\bibfnamefont {B.~A.}\ \bibnamefont
  {Bernevig}}, \ and\ \bibinfo {author} {\bibfnamefont {T.~L.}\ \bibnamefont
  {Hughes}},\ }\bibfield  {title} {\enquote {\bibinfo {title} {{Quantized
  electric multipole insulators}},}\ }\href@noop {} {\bibfield  {journal}
  {\bibinfo  {journal} {Science}\ }\textbf {\bibinfo {volume} {357}},\ \bibinfo
  {pages} {61--66} (\bibinfo {year} {2017})}\BibitemShut {NoStop}%
\bibitem [{\citenamefont {Serra-Garcia}\ \emph {et~al.}(2018)\citenamefont
  {Serra-Garcia}, \citenamefont {Peri}, \citenamefont {S\"{u}sstrunk},
  \citenamefont {Bilal}, \citenamefont {Larsen}, \citenamefont {Villanueva},\
  and\ \citenamefont {Huber}}]{18Nature_HOTI_Photonic}%
  \BibitemOpen
  \bibfield  {author} {\bibinfo {author} {\bibfnamefont {M.}~\bibnamefont
  {Serra-Garcia}}, \bibinfo {author} {\bibfnamefont {V.}~\bibnamefont {Peri}},
  \bibinfo {author} {\bibfnamefont {R.}~\bibnamefont {S\"{u}sstrunk}}, \bibinfo
  {author} {\bibfnamefont {O.~R.}\ \bibnamefont {Bilal}}, \bibinfo {author}
  {\bibfnamefont {T.}~\bibnamefont {Larsen}}, \bibinfo {author} {\bibfnamefont
  {L.~G.}\ \bibnamefont {Villanueva}}, \ and\ \bibinfo {author} {\bibfnamefont
  {S.~D.}\ \bibnamefont {Huber}},\ }\bibfield  {title} {\enquote {\bibinfo
  {title} {{Observation of a phononic quadrupole topological insulator}},}\
  }\href@noop {} {\bibfield  {journal} {\bibinfo  {journal} {Nature}\ }\textbf
  {\bibinfo {volume} {555}},\ \bibinfo {pages} {342--345} (\bibinfo {year}
  {2018})}\BibitemShut {NoStop}%
\bibitem [{\citenamefont {Klein}\ \emph {et~al.}(2006)\citenamefont {Klein},
  \citenamefont {Enkrich}, \citenamefont {Wegener},\ and\ \citenamefont
  {Linden}}]{06Science_MagneticMeta}%
  \BibitemOpen
  \bibfield  {author} {\bibinfo {author} {\bibfnamefont {M.~W.}\ \bibnamefont
  {Klein}}, \bibinfo {author} {\bibfnamefont {C.}~\bibnamefont {Enkrich}},
  \bibinfo {author} {\bibfnamefont {M.}~\bibnamefont {Wegener}}, \ and\
  \bibinfo {author} {\bibfnamefont {S.}~\bibnamefont {Linden}},\ }\bibfield
  {title} {\enquote {\bibinfo {title} {{Second-harmonic generation from
  magnetic metamaterials}},}\ }\href@noop {} {\bibfield  {journal} {\bibinfo
  {journal} {Science}\ }\textbf {\bibinfo {volume} {313}},\ \bibinfo {pages}
  {502--504} (\bibinfo {year} {2006})}\BibitemShut {NoStop}%
\bibitem [{\citenamefont {Krasnok}, \citenamefont {Tymchenko},\ and\
  \citenamefont {Al\`u}(2018)}]{18MaterToday_metasurface}%
  \BibitemOpen
  \bibfield  {author} {\bibinfo {author} {\bibfnamefont {A.}~\bibnamefont
  {Krasnok}}, \bibinfo {author} {\bibfnamefont {M.}~\bibnamefont {Tymchenko}},
  \ and\ \bibinfo {author} {\bibfnamefont {A.}~\bibnamefont {Al\`u}},\
  }\bibfield  {title} {\enquote {\bibinfo {title} {{Nonlinear metasurfaces: a
  paradigm shift in nonlinear optics}},}\ }\href@noop {} {\bibfield  {journal}
  {\bibinfo  {journal} {Mater. Today}\ }\textbf {\bibinfo {volume} {21}},\
  \bibinfo {pages} {8--21} (\bibinfo {year} {2018})}\BibitemShut {NoStop}%
\bibitem [{\citenamefont {Collins}\ \emph {et~al.}(2018)\citenamefont
  {Collins}, \citenamefont {Hooper}, \citenamefont {Mark}, \citenamefont
  {Kuppe},\ and\ \citenamefont {Valev}}]{18NanoLett_SolelyChiral}%
  \BibitemOpen
  \bibfield  {author} {\bibinfo {author} {\bibfnamefont {J.~T.}\ \bibnamefont
  {Collins}}, \bibinfo {author} {\bibfnamefont {D.~C.}\ \bibnamefont {Hooper}},
  \bibinfo {author} {\bibfnamefont {A.~G.}\ \bibnamefont {Mark}}, \bibinfo
  {author} {\bibfnamefont {C.}~\bibnamefont {Kuppe}}, \ and\ \bibinfo {author}
  {\bibfnamefont {V.~K.}\ \bibnamefont {Valev}},\ }\bibfield  {title} {\enquote
  {\bibinfo {title} {{Second-harmonic generation optical rotation solely
  attributable to chirality in plasmonic metasurfaces}},}\ }\href@noop {}
  {\bibfield  {journal} {\bibinfo  {journal} {ACS Nano}\ }\textbf {\bibinfo
  {volume} {12}},\ \bibinfo {pages} {5445--5451} (\bibinfo {year}
  {2018})}\BibitemShut {NoStop}%
\bibitem [{\citenamefont {Yamada}\ \emph {et~al.}(2004)\citenamefont {Yamada},
  \citenamefont {Ogawa}, \citenamefont {Ishii}, \citenamefont {Sato},
  \citenamefont {Kawasaki}, \citenamefont {Akoh},\ and\ \citenamefont
  {Tokura}}]{04Science_InterfaceMagnetism}%
  \BibitemOpen
  \bibfield  {author} {\bibinfo {author} {\bibfnamefont {H.}~\bibnamefont
  {Yamada}}, \bibinfo {author} {\bibfnamefont {Y.}~\bibnamefont {Ogawa}},
  \bibinfo {author} {\bibfnamefont {Y.}~\bibnamefont {Ishii}}, \bibinfo
  {author} {\bibfnamefont {H.}~\bibnamefont {Sato}}, \bibinfo {author}
  {\bibfnamefont {M.}~\bibnamefont {Kawasaki}}, \bibinfo {author}
  {\bibfnamefont {H.}~\bibnamefont {Akoh}}, \ and\ \bibinfo {author}
  {\bibfnamefont {Y.}~\bibnamefont {Tokura}},\ }\bibfield  {title} {\enquote
  {\bibinfo {title} {{Engineered interface of magnetic oxides}},}\ }\href@noop
  {} {\bibfield  {journal} {\bibinfo  {journal} {Science}\ }\textbf {\bibinfo
  {volume} {305}},\ \bibinfo {pages} {646--648} (\bibinfo {year}
  {2004})}\BibitemShut {NoStop}%
\bibitem [{\citenamefont {Ogawa}\ \emph {et~al.}(2009)\citenamefont {Ogawa},
  \citenamefont {Miyano}, \citenamefont {Hosoda}, \citenamefont {Higuchi},
  \citenamefont {Bell}, \citenamefont {Hikita},\ and\ \citenamefont
  {Hwang}}]{09PRB_Ogawa_LAOSTO}%
  \BibitemOpen
  \bibfield  {author} {\bibinfo {author} {\bibfnamefont {N.}~\bibnamefont
  {Ogawa}}, \bibinfo {author} {\bibfnamefont {K.}~\bibnamefont {Miyano}},
  \bibinfo {author} {\bibfnamefont {M.}~\bibnamefont {Hosoda}}, \bibinfo
  {author} {\bibfnamefont {T.}~\bibnamefont {Higuchi}}, \bibinfo {author}
  {\bibfnamefont {C.}~\bibnamefont {Bell}}, \bibinfo {author} {\bibfnamefont
  {Y.}~\bibnamefont {Hikita}}, \ and\ \bibinfo {author} {\bibfnamefont {H.~Y.}\
  \bibnamefont {Hwang}},\ }\bibfield  {title} {\enquote {\bibinfo {title}
  {{Enhanced lattice polarization in ${\text{SrTiO}}_{3}/{\text{LaAlO}}_{3}$
  superlattices measured using optical second-harmonic generation}},}\
  }\href@noop {} {\bibfield  {journal} {\bibinfo  {journal} {Phys. Rev. B}\
  }\textbf {\bibinfo {volume} {80}},\ \bibinfo {pages} {081106} (\bibinfo
  {year} {2009})}\BibitemShut {NoStop}%
\bibitem [{\citenamefont {Seyler}\ \emph {et~al.}(2015)\citenamefont {Seyler},
  \citenamefont {Schaibley}, \citenamefont {Gong}, \citenamefont {Rivera},
  \citenamefont {Jones}, \citenamefont {Wu}, \citenamefont {Yan}, \citenamefont
  {Mandrus}, \citenamefont {Yao},\ and\ \citenamefont
  {Xu}}]{15NNano_WSe2_SHG_Electric_ZDXu}%
  \BibitemOpen
  \bibfield  {author} {\bibinfo {author} {\bibfnamefont {K.~L.}\ \bibnamefont
  {Seyler}}, \bibinfo {author} {\bibfnamefont {J.~R.}\ \bibnamefont
  {Schaibley}}, \bibinfo {author} {\bibfnamefont {P.}~\bibnamefont {Gong}},
  \bibinfo {author} {\bibfnamefont {P.}~\bibnamefont {Rivera}}, \bibinfo
  {author} {\bibfnamefont {A.~M.}\ \bibnamefont {Jones}}, \bibinfo {author}
  {\bibfnamefont {S.}~\bibnamefont {Wu}}, \bibinfo {author} {\bibfnamefont
  {J.}~\bibnamefont {Yan}}, \bibinfo {author} {\bibfnamefont {D.~G.}\
  \bibnamefont {Mandrus}}, \bibinfo {author} {\bibfnamefont {W.}~\bibnamefont
  {Yao}}, \ and\ \bibinfo {author} {\bibfnamefont {X.}~\bibnamefont {Xu}},\
  }\bibfield  {title} {\enquote {\bibinfo {title} {{Electrical control of
  second-harmonic generation in a WSe$_2$ monolayer transistor}},}\ }\href@noop
  {} {\bibfield  {journal} {\bibinfo  {journal} {Nature Nanotech.}\ }\textbf
  {\bibinfo {volume} {10}},\ \bibinfo {pages} {407--411} (\bibinfo {year}
  {2015})}\BibitemShut {NoStop}%
\bibitem [{\citenamefont {Wu}\ \emph {et~al.}(2017)\citenamefont {Wu},
  \citenamefont {Patankar}, \citenamefont {Morimoto}, \citenamefont {Nair},
  \citenamefont {Thewalt}, \citenamefont {Little}, \citenamefont {Analytis},
  \citenamefont {Moore},\ and\ \citenamefont
  {Orenstein}}]{17NPhys_TaAs_SHG_LuOrenstein}%
  \BibitemOpen
  \bibfield  {author} {\bibinfo {author} {\bibfnamefont {L.}~\bibnamefont
  {Wu}}, \bibinfo {author} {\bibfnamefont {S.}~\bibnamefont {Patankar}},
  \bibinfo {author} {\bibfnamefont {T.}~\bibnamefont {Morimoto}}, \bibinfo
  {author} {\bibfnamefont {N.~L.}\ \bibnamefont {Nair}}, \bibinfo {author}
  {\bibfnamefont {E.}~\bibnamefont {Thewalt}}, \bibinfo {author} {\bibfnamefont
  {A.}~\bibnamefont {Little}}, \bibinfo {author} {\bibfnamefont {J.~G.}\
  \bibnamefont {Analytis}}, \bibinfo {author} {\bibfnamefont {J.~E.}\
  \bibnamefont {Moore}}, \ and\ \bibinfo {author} {\bibfnamefont
  {J.}~\bibnamefont {Orenstein}},\ }\bibfield  {title} {\enquote {\bibinfo
  {title} {{Giant anisotropic nonlinear optical response in transition metal
  monopnictide Weyl semimetals}},}\ }\href@noop {} {\bibfield  {journal}
  {\bibinfo  {journal} {Nature Phys.}\ }\textbf {\bibinfo {volume} {13}},\
  \bibinfo {pages} {350--355} (\bibinfo {year} {2017})}\BibitemShut {NoStop}%
\bibitem [{\citenamefont {Zhang}\ \emph {et~al.}(2019)\citenamefont {Zhang},
  \citenamefont {Wang}, \citenamefont {Li}, \citenamefont {Shi}, \citenamefont
  {Wu}, \citenamefont {Lin}, \citenamefont {Zhang}, \citenamefont {Liu},
  \citenamefont {Liu}, \citenamefont {Wang}, \citenamefont {Dong},\ and\
  \citenamefont {Wang}}]{19PRX_MoTe_SHG_NLWang}%
  \BibitemOpen
  \bibfield  {author} {\bibinfo {author} {\bibfnamefont {M.~Y.}\ \bibnamefont
  {Zhang}}, \bibinfo {author} {\bibfnamefont {Z.~X.}\ \bibnamefont {Wang}},
  \bibinfo {author} {\bibfnamefont {Y.~N.}\ \bibnamefont {Li}}, \bibinfo
  {author} {\bibfnamefont {L.~Y.}\ \bibnamefont {Shi}}, \bibinfo {author}
  {\bibfnamefont {D.}~\bibnamefont {Wu}}, \bibinfo {author} {\bibfnamefont
  {T.}~\bibnamefont {Lin}}, \bibinfo {author} {\bibfnamefont {S.~J.}\
  \bibnamefont {Zhang}}, \bibinfo {author} {\bibfnamefont {Y.~Q.}\ \bibnamefont
  {Liu}}, \bibinfo {author} {\bibfnamefont {Q.~M.}\ \bibnamefont {Liu}},
  \bibinfo {author} {\bibfnamefont {J.}~\bibnamefont {Wang}}, \bibinfo {author}
  {\bibfnamefont {T.}~\bibnamefont {Dong}}, \ and\ \bibinfo {author}
  {\bibfnamefont {N.~L.}\ \bibnamefont {Wang}},\ }\bibfield  {title} {\enquote
  {\bibinfo {title} {{Light-induced subpicosecond lattice symmetry switch in
  ${\mathrm{MoTe}}_{2}$}},}\ }\href@noop {} {\bibfield  {journal} {\bibinfo
  {journal} {Phys. Rev. X}\ }\textbf {\bibinfo {volume} {9}},\ \bibinfo {pages}
  {021036} (\bibinfo {year} {2019})}\BibitemShut {NoStop}%
\bibitem [{\citenamefont {Tom}, \citenamefont {Heinz},\ and\ \citenamefont
  {Shen}(1983)}]{83PRL_Si_VacuumRot}%
  \BibitemOpen
  \bibfield  {author} {\bibinfo {author} {\bibfnamefont {H.~W.~K.}\
  \bibnamefont {Tom}}, \bibinfo {author} {\bibfnamefont {T.~F.}\ \bibnamefont
  {Heinz}}, \ and\ \bibinfo {author} {\bibfnamefont {Y.~R.}\ \bibnamefont
  {Shen}},\ }\bibfield  {title} {\enquote {\bibinfo {title} {{Second-harmonic
  reflection from silicon surfaces and its relation to structural symmetry}},}\
  }\href@noop {} {\bibfield  {journal} {\bibinfo  {journal} {Phys. Rev. Lett.}\
  }\textbf {\bibinfo {volume} {51}},\ \bibinfo {pages} {1983--1986} (\bibinfo
  {year} {1983})}\BibitemShut {NoStop}%
\bibitem [{\citenamefont {Yamada}\ and\ \citenamefont
  {Kimura}(1993)}]{93PRL_GaAs_VacuumRot}%
  \BibitemOpen
  \bibfield  {author} {\bibinfo {author} {\bibfnamefont {C.}~\bibnamefont
  {Yamada}}\ and\ \bibinfo {author} {\bibfnamefont {T.}~\bibnamefont
  {Kimura}},\ }\bibfield  {title} {\enquote {\bibinfo {title} {{Anisotropy in
  second-harmonic generation from reconstructed surfaces of GaAs}},}\
  }\href@noop {} {\bibfield  {journal} {\bibinfo  {journal} {Phys. Rev. Lett.}\
  }\textbf {\bibinfo {volume} {70}},\ \bibinfo {pages} {2344--2347} (\bibinfo
  {year} {1993})}\BibitemShut {NoStop}%
\bibitem [{\citenamefont {Sato}\ \emph {et~al.}(2000)\citenamefont {Sato},
  \citenamefont {Kodama}, \citenamefont {Miyamoto}, \citenamefont {Takanashi},
  \citenamefont {Fujimori},\ and\ \citenamefont
  {Rasing}}]{00JAP_Sato_Magnetic_Rotation}%
  \BibitemOpen
  \bibfield  {author} {\bibinfo {author} {\bibfnamefont {K.}~\bibnamefont
  {Sato}}, \bibinfo {author} {\bibfnamefont {A.}~\bibnamefont {Kodama}},
  \bibinfo {author} {\bibfnamefont {M.}~\bibnamefont {Miyamoto}}, \bibinfo
  {author} {\bibfnamefont {K.}~\bibnamefont {Takanashi}}, \bibinfo {author}
  {\bibfnamefont {H.}~\bibnamefont {Fujimori}}, \ and\ \bibinfo {author}
  {\bibfnamefont {T.}~\bibnamefont {Rasing}},\ }\bibfield  {title} {\enquote
  {\bibinfo {title} {{Nonlinear magneto-optical effect in Fe/Au superlattices
  modulated by noninteger atomic layers}},}\ }\href@noop {} {\bibfield
  {journal} {\bibinfo  {journal} {J. Appl. Phys.}\ }\textbf {\bibinfo {volume}
  {87}},\ \bibinfo {pages} {6785--6787} (\bibinfo {year} {2000})}\BibitemShut
  {NoStop}%
\bibitem [{\citenamefont {Valev}\ \emph {et~al.}(2010)\citenamefont {Valev},
  \citenamefont {Silhanek}, \citenamefont {Verellen}, \citenamefont {Gillijns},
  \citenamefont {Van~Dorpe}, \citenamefont {Aktsipetrov}, \citenamefont
  {Vandenbosch}, \citenamefont {Moshchalkov},\ and\ \citenamefont
  {Verbiest}}]{10PRL_GoldNano_SHG_Valev}%
  \BibitemOpen
  \bibfield  {author} {\bibinfo {author} {\bibfnamefont {V.~K.}\ \bibnamefont
  {Valev}}, \bibinfo {author} {\bibfnamefont {A.~V.}\ \bibnamefont {Silhanek}},
  \bibinfo {author} {\bibfnamefont {N.}~\bibnamefont {Verellen}}, \bibinfo
  {author} {\bibfnamefont {W.}~\bibnamefont {Gillijns}}, \bibinfo {author}
  {\bibfnamefont {P.}~\bibnamefont {Van~Dorpe}}, \bibinfo {author}
  {\bibfnamefont {O.~A.}\ \bibnamefont {Aktsipetrov}}, \bibinfo {author}
  {\bibfnamefont {G.~A.~E.}\ \bibnamefont {Vandenbosch}}, \bibinfo {author}
  {\bibfnamefont {V.~V.}\ \bibnamefont {Moshchalkov}}, \ and\ \bibinfo {author}
  {\bibfnamefont {T.}~\bibnamefont {Verbiest}},\ }\bibfield  {title} {\enquote
  {\bibinfo {title} {{Asymmetric optical second-harmonic generation from chiral
  $G$-shaped gold nanostructures}},}\ }\href@noop {} {\bibfield  {journal}
  {\bibinfo  {journal} {Phys. Rev. Lett.}\ }\textbf {\bibinfo {volume} {104}},\
  \bibinfo {pages} {127401} (\bibinfo {year} {2010})}\BibitemShut {NoStop}%
\bibitem [{\citenamefont {Torchinsky}\ \emph {et~al.}(2014)\citenamefont
  {Torchinsky}, \citenamefont {Chu}, \citenamefont {Qi}, \citenamefont {Cao},\
  and\ \citenamefont {Hsieh}}]{14RSI_Torchinsky}%
  \BibitemOpen
  \bibfield  {author} {\bibinfo {author} {\bibfnamefont {D.~H.}\ \bibnamefont
  {Torchinsky}}, \bibinfo {author} {\bibfnamefont {H.}~\bibnamefont {Chu}},
  \bibinfo {author} {\bibfnamefont {T.}~\bibnamefont {Qi}}, \bibinfo {author}
  {\bibfnamefont {G.}~\bibnamefont {Cao}}, \ and\ \bibinfo {author}
  {\bibfnamefont {D.}~\bibnamefont {Hsieh}},\ }\bibfield  {title} {\enquote
  {\bibinfo {title} {{A low temperature nonlinear optical rotational anisotropy
  spectrometer for the determination of crystallographic and electronic
  symmetries}},}\ }\href@noop {} {\bibfield  {journal} {\bibinfo  {journal}
  {Rev. Sci. Instrum.}\ }\textbf {\bibinfo {volume} {85}},\ \bibinfo {pages}
  {083102} (\bibinfo {year} {2014})}\BibitemShut {NoStop}%
\bibitem [{\citenamefont {Harter}\ \emph {et~al.}(2015)\citenamefont {Harter},
  \citenamefont {Niu}, \citenamefont {Woss},\ and\ \citenamefont
  {Hsieh}}]{15OptLett_Harter}%
  \BibitemOpen
  \bibfield  {author} {\bibinfo {author} {\bibfnamefont {J.~W.}\ \bibnamefont
  {Harter}}, \bibinfo {author} {\bibfnamefont {L.}~\bibnamefont {Niu}},
  \bibinfo {author} {\bibfnamefont {A.~J.}\ \bibnamefont {Woss}}, \ and\
  \bibinfo {author} {\bibfnamefont {D.}~\bibnamefont {Hsieh}},\ }\bibfield
  {title} {\enquote {\bibinfo {title} {{High-speed measurement of rotational
  anisotropy nonlinear optical harmonic generation using position-sensitive
  detection}},}\ }\href@noop {} {\bibfield  {journal} {\bibinfo  {journal}
  {Opt. Lett.}\ }\textbf {\bibinfo {volume} {40}},\ \bibinfo {pages}
  {4671--4674} (\bibinfo {year} {2015})}\BibitemShut {NoStop}%
\bibitem [{\citenamefont {Lu}, \citenamefont {Tran},\ and\ \citenamefont
  {Torchinsky}(2019)}]{19RSI_LuTorchinsky}%
  \BibitemOpen
  \bibfield  {author} {\bibinfo {author} {\bibfnamefont {B.}~\bibnamefont
  {Lu}}, \bibinfo {author} {\bibfnamefont {J.~D.}\ \bibnamefont {Tran}}, \ and\
  \bibinfo {author} {\bibfnamefont {D.~H.}\ \bibnamefont {Torchinsky}},\
  }\bibfield  {title} {\enquote {\bibinfo {title} {{Fast reflective optic-based
  rotational anisotropy nonlinear harmonic generation spectrometer}},}\
  }\href@noop {} {\bibfield  {journal} {\bibinfo  {journal} {Rev. Sci.
  Instrum.}\ }\textbf {\bibinfo {volume} {90}},\ \bibinfo {pages} {053102}
  (\bibinfo {year} {2019})}\BibitemShut {NoStop}%
\bibitem [{\citenamefont {Fichera}\ \emph {et~al.}(2020)\citenamefont
  {Fichera}, \citenamefont {Kogar}, \citenamefont {Ye}, \citenamefont
  {G\"okce}, \citenamefont {Zong}, \citenamefont {Checkelsky},\ and\
  \citenamefont {Gedik}}]{20PRB_RA_SHG_Gedik}%
  \BibitemOpen
  \bibfield  {author} {\bibinfo {author} {\bibfnamefont {B.~T.}\ \bibnamefont
  {Fichera}}, \bibinfo {author} {\bibfnamefont {A.}~\bibnamefont {Kogar}},
  \bibinfo {author} {\bibfnamefont {L.}~\bibnamefont {Ye}}, \bibinfo {author}
  {\bibfnamefont {B.}~\bibnamefont {G\"okce}}, \bibinfo {author} {\bibfnamefont
  {A.}~\bibnamefont {Zong}}, \bibinfo {author} {\bibfnamefont {J.~G.}\
  \bibnamefont {Checkelsky}}, \ and\ \bibinfo {author} {\bibfnamefont
  {N.}~\bibnamefont {Gedik}},\ }\bibfield  {title} {\enquote {\bibinfo {title}
  {{Second harmonic generation as a probe of broken mirror symmetry}},}\
  }\href@noop {} {\bibfield  {journal} {\bibinfo  {journal} {Phys. Rev. B}\
  }\textbf {\bibinfo {volume} {101}},\ \bibinfo {pages} {241106} (\bibinfo
  {year} {2020})}\BibitemShut {NoStop}%
\bibitem [{\citenamefont {Sheu}\ \emph {et~al.}(2014)\citenamefont {Sheu},
  \citenamefont {Trugman}, \citenamefont {Yan}, \citenamefont {Jia},
  \citenamefont {Taylor},\ and\ \citenamefont
  {Prasankumar}}]{14NCom_Ferro_TrSHG}%
  \BibitemOpen
  \bibfield  {author} {\bibinfo {author} {\bibfnamefont {Y.~M.}\ \bibnamefont
  {Sheu}}, \bibinfo {author} {\bibfnamefont {S.~A.}\ \bibnamefont {Trugman}},
  \bibinfo {author} {\bibfnamefont {L.}~\bibnamefont {Yan}}, \bibinfo {author}
  {\bibfnamefont {Q.~X.}\ \bibnamefont {Jia}}, \bibinfo {author} {\bibfnamefont
  {A.~J.}\ \bibnamefont {Taylor}}, \ and\ \bibinfo {author} {\bibfnamefont
  {R.~P.}\ \bibnamefont {Prasankumar}},\ }\bibfield  {title} {\enquote
  {\bibinfo {title} {{Using ultrashort optical pulses to couple ferroelectric
  and ferromagnetic order in an oxide heterostructure}},}\ }\href@noop {}
  {\bibfield  {journal} {\bibinfo  {journal} {Nature Commun.}\ }\textbf
  {\bibinfo {volume} {5}},\ \bibinfo {pages} {5832} (\bibinfo {year}
  {2014})}\BibitemShut {NoStop}%
\bibitem [{\citenamefont {Jang}\ \emph {et~al.}(2018)\citenamefont {Jang},
  \citenamefont {Dhakal}, \citenamefont {J}, \citenamefont {Yun}, \citenamefont
  {Shinde}, \citenamefont {Chen}, \citenamefont {Jeong}, \citenamefont {Lee},
  \citenamefont {Lee}, \citenamefont {Lee}, \citenamefont {Ahn},\ and\
  \citenamefont {Kim}}]{18AdvMater_TSHG_MoSe}%
  \BibitemOpen
  \bibfield  {author} {\bibinfo {author} {\bibfnamefont {H.}~\bibnamefont
  {Jang}}, \bibinfo {author} {\bibfnamefont {K.~P.}\ \bibnamefont {Dhakal}},
  \bibinfo {author} {\bibfnamefont {K.-I.}\ \bibnamefont {Joo}}, \bibinfo
  {author} {\bibfnamefont {W.~S.}\ \bibnamefont {Yun}}, \bibinfo {author}
  {\bibfnamefont {S.~M.}\ \bibnamefont {Shinde}}, \bibinfo {author}
  {\bibfnamefont {X.}~\bibnamefont {Chen}}, \bibinfo {author} {\bibfnamefont
  {S.~M.}\ \bibnamefont {Jeong}}, \bibinfo {author} {\bibfnamefont {S.~W.}\
  \bibnamefont {Lee}}, \bibinfo {author} {\bibfnamefont {Z.}~\bibnamefont
  {Lee}}, \bibinfo {author} {\bibfnamefont {J.~D.}\ \bibnamefont {Lee}},
  \bibinfo {author} {\bibfnamefont {J.-H.}\ \bibnamefont {Ahn}}, \ and\
  \bibinfo {author} {\bibfnamefont {H.}~\bibnamefont {Kim}},\ }\bibfield
  {title} {\enquote {\bibinfo {title} {{Transient SHG imaging on ultrafast
  carrier dynamics of MoS$_2$ nanosheets}},}\ }\href@noop {} {\bibfield
  {journal} {\bibinfo  {journal} {Adv. Mater.}\ }\textbf {\bibinfo {volume}
  {30}},\ \bibinfo {pages} {1705190} (\bibinfo {year} {2018})}\BibitemShut
  {NoStop}%
\bibitem [{\citenamefont {Fermann}\ and\ \citenamefont
  {Hartl}(2013)}]{13NPhoton_Rev_UltrafastFiber}%
  \BibitemOpen
  \bibfield  {author} {\bibinfo {author} {\bibfnamefont {M.~E.}\ \bibnamefont
  {Fermann}}\ and\ \bibinfo {author} {\bibfnamefont {I.}~\bibnamefont
  {Hartl}},\ }\bibfield  {title} {\enquote {\bibinfo {title} {{Ultrafast fibre
  lasers}},}\ }\href@noop {} {\bibfield  {journal} {\bibinfo  {journal} {Nat.
  Photon.}\ }\textbf {\bibinfo {volume} {7}},\ \bibinfo {pages} {868--874}
  (\bibinfo {year} {2013})}\BibitemShut {NoStop}%
\bibitem [{\citenamefont {Chong}\ \emph {et~al.}(2006)\citenamefont {Chong},
  \citenamefont {Buckley}, \citenamefont {Renninger},\ and\ \citenamefont
  {Wise}}]{06OptExp_ANDi}%
  \BibitemOpen
  \bibfield  {author} {\bibinfo {author} {\bibfnamefont {A.}~\bibnamefont
  {Chong}}, \bibinfo {author} {\bibfnamefont {J.}~\bibnamefont {Buckley}},
  \bibinfo {author} {\bibfnamefont {W.}~\bibnamefont {Renninger}}, \ and\
  \bibinfo {author} {\bibfnamefont {F.}~\bibnamefont {Wise}},\ }\bibfield
  {title} {\enquote {\bibinfo {title} {{All-normal-dispersion femtosecond fiber
  laser}},}\ }\href@noop {} {\bibfield  {journal} {\bibinfo  {journal} {Opt.
  Express}\ }\textbf {\bibinfo {volume} {14}},\ \bibinfo {pages} {10095--10100}
  (\bibinfo {year} {2006})}\BibitemShut {NoStop}%
\bibitem [{\citenamefont {Kumar}\ \emph {et~al.}(2013)\citenamefont {Kumar},
  \citenamefont {Najmaei}, \citenamefont {Cui}, \citenamefont {Ceballos},
  \citenamefont {Ajayan}, \citenamefont {Lou},\ and\ \citenamefont
  {Zhao}}]{13PRB_SHG_MoS2_Kumar}%
  \BibitemOpen
  \bibfield  {author} {\bibinfo {author} {\bibfnamefont {N.}~\bibnamefont
  {Kumar}}, \bibinfo {author} {\bibfnamefont {S.}~\bibnamefont {Najmaei}},
  \bibinfo {author} {\bibfnamefont {Q.}~\bibnamefont {Cui}}, \bibinfo {author}
  {\bibfnamefont {F.}~\bibnamefont {Ceballos}}, \bibinfo {author}
  {\bibfnamefont {P.~M.}\ \bibnamefont {Ajayan}}, \bibinfo {author}
  {\bibfnamefont {J.}~\bibnamefont {Lou}}, \ and\ \bibinfo {author}
  {\bibfnamefont {H.}~\bibnamefont {Zhao}},\ }\bibfield  {title} {\enquote
  {\bibinfo {title} {{Second harmonic microscopy of monolayer MoS${}_{2}$}},}\
  }\href@noop {} {\bibfield  {journal} {\bibinfo  {journal} {Phys. Rev. B}\
  }\textbf {\bibinfo {volume} {87}},\ \bibinfo {pages} {161403(R)} (\bibinfo
  {year} {2013})}\BibitemShut {NoStop}%
\bibitem [{\citenamefont {Malard}\ \emph {et~al.}(2013)\citenamefont {Malard},
  \citenamefont {Alencar}, \citenamefont {Barboza}, \citenamefont {Mak},\ and\
  \citenamefont {de~Paula}}]{13PRB_SHG_MoS2_Malard}%
  \BibitemOpen
  \bibfield  {author} {\bibinfo {author} {\bibfnamefont {L.~M.}\ \bibnamefont
  {Malard}}, \bibinfo {author} {\bibfnamefont {T.~V.}\ \bibnamefont {Alencar}},
  \bibinfo {author} {\bibfnamefont {A.~P.~M.}\ \bibnamefont {Barboza}},
  \bibinfo {author} {\bibfnamefont {K.~F.}\ \bibnamefont {Mak}}, \ and\
  \bibinfo {author} {\bibfnamefont {A.~M.}\ \bibnamefont {de~Paula}},\
  }\bibfield  {title} {\enquote {\bibinfo {title} {{Observation of intense
  second harmonic generation from MoS${}_{2}$ atomic crystals}},}\ }\href@noop
  {} {\bibfield  {journal} {\bibinfo  {journal} {Phys. Rev. B}\ }\textbf
  {\bibinfo {volume} {87}},\ \bibinfo {pages} {201401(R)} (\bibinfo {year}
  {2013})}\BibitemShut {NoStop}%
\bibitem [{\citenamefont {Li}\ \emph {et~al.}(2013)\citenamefont {Li},
  \citenamefont {Rao}, \citenamefont {Mak}, \citenamefont {You}, \citenamefont
  {Wang}, \citenamefont {Dean},\ and\ \citenamefont
  {Heinz}}]{13NanoLett_SHG_MoS2_Heinz}%
  \BibitemOpen
  \bibfield  {author} {\bibinfo {author} {\bibfnamefont {Y.}~\bibnamefont
  {Li}}, \bibinfo {author} {\bibfnamefont {Y.}~\bibnamefont {Rao}}, \bibinfo
  {author} {\bibfnamefont {K.~F.}\ \bibnamefont {Mak}}, \bibinfo {author}
  {\bibfnamefont {Y.}~\bibnamefont {You}}, \bibinfo {author} {\bibfnamefont
  {S.}~\bibnamefont {Wang}}, \bibinfo {author} {\bibfnamefont {C.~R.}\
  \bibnamefont {Dean}}, \ and\ \bibinfo {author} {\bibfnamefont {T.~F.}\
  \bibnamefont {Heinz}},\ }\bibfield  {title} {\enquote {\bibinfo {title}
  {{Probing symmetry properties of few-layer MoS$_2$ and h-BN by optical
  second-harmonic generation}},}\ }\href@noop {} {\bibfield  {journal}
  {\bibinfo  {journal} {Nano Lett.}\ }\textbf {\bibinfo {volume} {13}},\
  \bibinfo {pages} {3329--3333} (\bibinfo {year} {2013})}\BibitemShut {NoStop}%
\bibitem [{\citenamefont {Janisch}\ \emph {et~al.}(2014)\citenamefont
  {Janisch}, \citenamefont {Wang}, \citenamefont {Ma}, \citenamefont {Mehta},
  \citenamefont {El\'{i}as}, \citenamefont {Perea-L\'{o}pez}, \citenamefont
  {Terrones}, \citenamefont {Crespi},\ and\ \citenamefont
  {Liu}}]{14SciRep_SHG_WS2}%
  \BibitemOpen
  \bibfield  {author} {\bibinfo {author} {\bibfnamefont {C.}~\bibnamefont
  {Janisch}}, \bibinfo {author} {\bibfnamefont {Y.}~\bibnamefont {Wang}},
  \bibinfo {author} {\bibfnamefont {D.}~\bibnamefont {Ma}}, \bibinfo {author}
  {\bibfnamefont {N.}~\bibnamefont {Mehta}}, \bibinfo {author} {\bibfnamefont
  {A.~L.}\ \bibnamefont {El\'{i}as}}, \bibinfo {author} {\bibfnamefont
  {N.}~\bibnamefont {Perea-L\'{o}pez}}, \bibinfo {author} {\bibfnamefont
  {M.}~\bibnamefont {Terrones}}, \bibinfo {author} {\bibfnamefont
  {V.}~\bibnamefont {Crespi}}, \ and\ \bibinfo {author} {\bibfnamefont
  {Z.}~\bibnamefont {Liu}},\ }\bibfield  {title} {\enquote {\bibinfo {title}
  {{Extraordinary second harmonic generation in tungsten disulfide
  monolayers}},}\ }\href@noop {} {\bibfield  {journal} {\bibinfo  {journal}
  {Sci. Rep.}\ }\textbf {\bibinfo {volume} {4}},\ \bibinfo {pages} {5530}
  (\bibinfo {year} {2014})}\BibitemShut {NoStop}%
\bibitem [{\citenamefont {Ribeiro-Soares}\ \emph {et~al.}(2015)\citenamefont
  {Ribeiro-Soares}, \citenamefont {Janisch}, \citenamefont {Liu}, \citenamefont
  {El\'{i}as}, \citenamefont {Dresselhaus}, \citenamefont {Terrones},
  \citenamefont {Can\c{c}ado},\ and\ \citenamefont
  {Jorio}}]{15TwoDMater_SHG_WSe2}%
  \BibitemOpen
  \bibfield  {author} {\bibinfo {author} {\bibfnamefont {J.}~\bibnamefont
  {Ribeiro-Soares}}, \bibinfo {author} {\bibfnamefont {C.}~\bibnamefont
  {Janisch}}, \bibinfo {author} {\bibfnamefont {Z.}~\bibnamefont {Liu}},
  \bibinfo {author} {\bibfnamefont {A.~L.}\ \bibnamefont {El\'{i}as}}, \bibinfo
  {author} {\bibfnamefont {M.~S.}\ \bibnamefont {Dresselhaus}}, \bibinfo
  {author} {\bibfnamefont {M.}~\bibnamefont {Terrones}}, \bibinfo {author}
  {\bibfnamefont {L.~G.}\ \bibnamefont {Can\c{c}ado}}, \ and\ \bibinfo {author}
  {\bibfnamefont {A.}~\bibnamefont {Jorio}},\ }\bibfield  {title} {\enquote
  {\bibinfo {title} {{Second harmonic generation in WSe$_2$}},}\ }\href@noop {}
  {\bibfield  {journal} {\bibinfo  {journal} {2D Mater.}\ }\textbf {\bibinfo
  {volume} {2}},\ \bibinfo {pages} {045015} (\bibinfo {year}
  {2015})}\BibitemShut {NoStop}%
\bibitem [{\citenamefont {Rosa}\ \emph {et~al.}(2018)\citenamefont {Rosa},
  \citenamefont {Ho}, \citenamefont {Verzhbitskiy}, \citenamefont {Rodrigues},
  \citenamefont {Taniguchi}, \citenamefont {Watanabe}, \citenamefont {Eda},
  \citenamefont {Pereira},\ and\ \citenamefont {Gomes}}]{18SciRep_SHG_WSe2}%
  \BibitemOpen
  \bibfield  {author} {\bibinfo {author} {\bibfnamefont {H.~G.}\ \bibnamefont
  {Rosa}}, \bibinfo {author} {\bibfnamefont {Y.~W.}\ \bibnamefont {Ho}},
  \bibinfo {author} {\bibfnamefont {I.}~\bibnamefont {Verzhbitskiy}}, \bibinfo
  {author} {\bibfnamefont {M.~J. F.~L.}\ \bibnamefont {Rodrigues}}, \bibinfo
  {author} {\bibfnamefont {T.}~\bibnamefont {Taniguchi}}, \bibinfo {author}
  {\bibfnamefont {K.}~\bibnamefont {Watanabe}}, \bibinfo {author}
  {\bibfnamefont {G.}~\bibnamefont {Eda}}, \bibinfo {author} {\bibfnamefont
  {V.~M.}\ \bibnamefont {Pereira}}, \ and\ \bibinfo {author} {\bibfnamefont
  {J.~C.~V.}\ \bibnamefont {Gomes}},\ }\bibfield  {title} {\enquote {\bibinfo
  {title} {{Characterization of the second- and third-harmonic optical
  susceptibilities of atomically thin tungsten diselenide}},}\ }\href@noop {}
  {\bibfield  {journal} {\bibinfo  {journal} {Sci. Rep.}\ }\textbf {\bibinfo
  {volume} {8}},\ \bibinfo {pages} {10035} (\bibinfo {year}
  {2018})}\BibitemShut {NoStop}%
\bibitem [{\citenamefont {Jatirian-Foltides}\ \emph {et~al.}(2016)\citenamefont
  {Jatirian-Foltides}, \citenamefont {Escobedo-Alatorre}, \citenamefont
  {M\'{a}rquez-Aguilar}, \citenamefont {Hardhienata}, \citenamefont {Hingerl},\
  and\ \citenamefont {Alejo-Molina}}]{16RevMex_Chi2}%
  \BibitemOpen
  \bibfield  {author} {\bibinfo {author} {\bibfnamefont {E.~S.}\ \bibnamefont
  {Jatirian-Foltides}}, \bibinfo {author} {\bibfnamefont {J.~J.}\ \bibnamefont
  {Escobedo-Alatorre}}, \bibinfo {author} {\bibfnamefont {P.~A.}\ \bibnamefont
  {M\'{a}rquez-Aguilar}}, \bibinfo {author} {\bibfnamefont {H.}~\bibnamefont
  {Hardhienata}}, \bibinfo {author} {\bibfnamefont {K.}~\bibnamefont
  {Hingerl}}, \ and\ \bibinfo {author} {\bibfnamefont {A.}~\bibnamefont
  {Alejo-Molina}},\ }\bibfield  {title} {\enquote {\bibinfo {title} {{About the
  calculation of the second-order $\chi^{(2)}$ susceptibility tensorial
  elements for crystals using group theory}},}\ }\href@noop {} {\bibfield
  {journal} {\bibinfo  {journal} {Rev. Mex. Fis. E}\ }\textbf {\bibinfo
  {volume} {62}},\ \bibinfo {pages} {5--13} (\bibinfo {year}
  {2016})}\BibitemShut {NoStop}%
\bibitem [{\citenamefont {Hardhienata}\ \emph {et~al.}(2016)\citenamefont
  {Hardhienata}, \citenamefont {Alejo-Molina}, \citenamefont {Reitb\"{o}ck},
  \citenamefont {Prylepa}, \citenamefont {Stifter},\ and\ \citenamefont
  {Hingerl}}]{16JOptSocAmB_SHG_ZincBlend}%
  \BibitemOpen
  \bibfield  {author} {\bibinfo {author} {\bibfnamefont {H.}~\bibnamefont
  {Hardhienata}}, \bibinfo {author} {\bibfnamefont {A.}~\bibnamefont
  {Alejo-Molina}}, \bibinfo {author} {\bibfnamefont {C.}~\bibnamefont
  {Reitb\"{o}ck}}, \bibinfo {author} {\bibfnamefont {A.}~\bibnamefont
  {Prylepa}}, \bibinfo {author} {\bibfnamefont {D.}~\bibnamefont {Stifter}}, \
  and\ \bibinfo {author} {\bibfnamefont {K.}~\bibnamefont {Hingerl}},\
  }\bibfield  {title} {\enquote {\bibinfo {title} {{Bulk dipolar contribution
  to second-harmonic generation in zincblende}},}\ }\href@noop {} {\bibfield
  {journal} {\bibinfo  {journal} {J. Opt. Soc. Am. B}\ }\textbf {\bibinfo
  {volume} {33}},\ \bibinfo {pages} {195--201} (\bibinfo {year}
  {2016})}\BibitemShut {NoStop}%
\end{thebibliography}

%

\end{document}